\pdfoutput=1
\documentclass[cernpreprint, atlasdraft=false, texlive=2023, UKenglish, texmf, orcidlogo]{atlasdoc}
 
\usepackage[subcaption]{atlaspackage}
\usepackage{atlasbiblatex}
 
\usepackage[BSM]{atlasphysics}

\graphicspath{{logos/}{figures/}}
 
\usepackage{ANA-SUSY-2020-27-PAPER-defs}
 
\usepackage{multirow}
\usepackage{tabularx}
\usepackage{longtable}
\usepackage{ltablex}

\AtlasTitle{Search for pair production of squarks or gluinos
decaying via sleptons or weak bosons
in final states with two same-sign or three leptons with the ATLAS detector}

\AtlasAbstract{
A search for pair production of squarks or gluinos decaying via sleptons or weak bosons is reported.
The search  targets a final state with exactly two leptons with same-sign electric charge or at least three leptons without any charge requirement.
The analysed data set corresponds to an integrated luminosity of 139~fb$^{-1}$ of  proton--proton collisions  collected at a centre-of-mass energy of 13~TeV with the ATLAS detector at the LHC.
Multiple signal regions are defined, targeting several SUSY simplified models yielding the desired final states.
A single control region is used to constrain the normalisation of the $WZ$\,+\,jets background.
No significant excess of events over the Standard Model expectation is observed.
The results are interpreted in the context of several supersymmetric models featuring R-parity conservation or R-parity violation, yielding exclusion limits surpassing those from previous searches.
In models considering gluino (squark) pair production, gluino (squark) masses up to 2.2~(1.7)~TeV are excluded at 95\% confidence level.
}

\PreprintIdNumber{CERN-EP-2023-123}

\AtlasJournal{JHEP}
\AtlasJournalRef{\JHEP 02 (2024) 107}
\AtlasDOI{10.1007/JHEP02(2024)107}

\hypersetup{pdftitle={ATLAS document},pdfauthor={The ATLAS Collaboration}}
 
\begin{document}
 
\maketitle
 
\tableofcontents
 
\section{Introduction}

The ATLAS experiment~\cite{PERF-2007-01}
at the Large Hadron Collider (LHC)~\cite{Evans:2008zzb}
probes the electroweak sector of the Standard Model (SM) at the \TeV\ scale.
The presence of prompt electrons or muons (collectively referred to as leptons) in reconstructed events
provides one of the main experimental signatures to isolate processes
mediated by electroweak, scalar or exotic couplings,
from the large QCD multijet background produced in proton--proton ($pp$) collisions.
The production of pairs of leptons with the same electric charge
(also referred to as \enquote{same-sign leptons})
is particularly rare in the SM,
with an inclusive cross section of about \SI{1}{pb}~\cite{Alwall:2014hca}
for centre-of-mass energies around \SI{13}{\TeV}.
In contrast, it may occur frequently in scenarios with physics beyond the SM (BSM physics)~\cite{
DeSimone:2012fs,Fuks:2019clu,Barnett:1993ea},
and therefore searches for anomalous production of same-sign leptons
have been integral parts of the LHC and Tevatron experimental programmes~\cite{
EXOT-2012-13,CMS-SUS-11-020,PhysRevLett.93.061802}.
 
Extensions of the SM introducing
invariance under supersymmetric transformations~\cite{
Golfand:1971iw,Volkov:1973ix,Wess:1974tw,Wess:1974jb,Ferrara:1974pu,Salam:1974ig} (SUSY)
provide many such possibilities.
Even minimal realisations such as the
Minimal Supersymmetric Standard Model (MSSM)~\cite{Fayet:1976et,Fayet:1977yc}
contain SUSY partners for all SM fields
as well as members of an extended Higgs sector~\cite{Fayet:1974pd}
that may decay in complex cascades involving leptons.
In the MSSM, SUSY transformations relate
each of the Weyl components of a fundamental SM chiral fermion $f$
to a new scalar field, $\tilde f_\text{L}$ or $\tilde f_\text{R}$,
with identical gauge charges.
The quarks and leptons thus lead to 12 physical squarks (\squark) and 9 sleptons (\slepton, \snu).
Gluinos \gluino are the spin-$1/2$ Majorana fermionic partners of SM gluons.
The partners of electroweak and Higgs bosons mix to form spin-$1/2$ mass eigenstates
referred to as neutralinos $\tilde\chi^0_i$ ($i=1,\ldots 4$, ordered by increasing mass)
for the neutral ones,
and as charginos $\tilde\chi_j^\pm$ ($j=1,2$) for the others.
Depending on the dominant components of the admixtures,
they might be described as bino-, wino- or higgsino-like,
with important consequences for the mass spectrum and main decay channels~\cite{Canepa:2020ntc}.
 
By assuming an ad hoc discrete symmetry, the R-parity~\cite{Farrar:1978xj},
the lightest SUSY particle (LSP) becomes stable
and may contribute to dark matter~\cite{Goldberg:1983nd,Ellis:1983ew}.
In many models the LSP is the lightest neutralino \ninoone,
which interacts weakly and thus leaves missing transverse momentum as its
characteristic signature in the detector.
Other phenomenological consequences in R-parity-conserving (RPC) models
include SUSY partners always being produced in pairs,
with cascade decays into a final state of LSPs and SM particles.
On the other hand,
R-parity-violating (RPV) models~\cite{Girardello:1981wz}
may allow non-conservation of baryon number or lepton number,
potentially necessary for grand unification~\cite{Dreiner:1997uz}
or neutrino flavour mixing.
Detector signatures in such models are highly variable since they
depend on the nature and strength of the non-zero RPV couplings,
and can significantly enhance the production of multilepton final states at the LHC.
 
A search for pair production of gluinos or squarks
with the ATLAS experiment is presented in the following.
Different types of cascade decays are considered, arising in either RPC or RPV SUSY scenarios,
which lead to final states with either two same-sign leptons or three leptons,
several jets and, in many cases, missing transverse momentum.
The analysis makes use of the full set of data collected during
\RunTwo of the LHC.
The results complement and improve upon those from an earlier search~\cite{SUSY-2018-09}
performed on the same data set,
yielding increased sensitivity for two benchmark scenarios
while also providing new tailored search regions for processes or decay modes
not considered in Ref.~\cite{SUSY-2018-09},
such as the production of first- or second-generation squarks.
A search with a similar purpose was performed by the CMS experiment~\cite{CMS-SUS-19-008}.
Models with squark production were probed previously by ATLAS during \RunOne~\cite{SUSY-2014-06}.
 
The paper is organised as follows.
An overview of the ATLAS detector is provided in Section~\ref{sec:detector},
followed in Section~\ref{sec:signals}
by descriptions of the different SUSY processes of relevance.
Details of the recorded data used for the analysis,
and of the simulated Monte Carlo (MC) samples, are given in Section~\ref{sec:samples},
while the reconstruction of different types of high-level objects from those inputs
is described in Section~\ref{sec:objects}.
The definitions of several search regions used to look for the chosen SUSY processes
are motivated in Section~\ref{sec:strategy}.
The estimation of SM backgrounds is described in Section~\ref{sec:background},
followed by a summary of the sources of systematic uncertainty
affecting background or signal predictions in Section~\ref{sec:systematics}.
The various estimates are compared with observations in control and signal regions
by using a coherent statistical framework described in Section~\ref{sec:statistical},
providing the results listed in Section~\ref{sec:result}
and used to perform signal-strength hypothesis tests for
the various SUSY benchmarks.
Concluding remarks are provided in Section~\ref{sec:conclusion}.


\section{ATLAS detector}
\label{sec:detector}

The ATLAS detector~\cite{PERF-2007-01} at the LHC covers nearly the entire solid angle around the collision point.\footnote{\AtlasCoordFootnote}
It consists of an inner tracking detector surrounded by a thin superconducting solenoid, electromagnetic and hadron calorimeters,
and a muon spectrometer incorporating three large superconducting air-core toroidal magnets.
 
The inner-detector system (ID) is immersed in a \qty{2}{\tesla} axial magnetic field
and provides charged-particle tracking in the range \(|\eta| < 2.5\).
The high-granularity silicon pixel detector covers the vertex region and typically provides four measurements per track,
the first hit normally being in the insertable B-layer installed before \RunTwo~\cite{ATLAS-TDR-19,PIX-2018-001}.
It is followed by the silicon microstrip tracker, which usually provides eight measurements per track.
These silicon detectors are complemented by the transition radiation tracker (TRT),
which enables radially extended track reconstruction up to \(|\eta| = 2.0\).
The TRT also provides electron identification information
based on the fraction of hits (typically 30 in total) above a higher energy-deposit threshold corresponding to transition radiation.
 
The calorimeter system covers the pseudorapidity range \(|\eta| < 4.9\).
Within the region \(|\eta|< 3.2\), electromagnetic calorimetry is provided by barrel and
endcap high-granularity lead/liquid-argon (LAr) calorimeters,
with an additional thin LAr presampler covering \(|\eta| < 1.8\)
to correct for energy loss in material upstream of the calorimeters.
Hadron calorimetry is provided by the steel/scintillator-tile calorimeter,
segmented into three barrel structures within \(|\eta| < 1.7\), and two copper/LAr hadron endcap calorimeters.
The solid angle coverage is completed with forward copper/LAr and tungsten/LAr calorimeter modules
optimised for electromagnetic and hadronic energy measurements respectively.
 
The muon spectrometer (MS) comprises separate trigger and
high-precision tracking chambers measuring the deflection of muons in a magnetic field generated by the superconducting air-core toroidal magnets.
The field integral of the toroids ranges between \num{2.0} and \qty{6.0}{\tesla\metre}
across most of the detector.
Three layers of precision chambers, each consisting of layers of monitored drift tubes, cover the region \(|\eta| < 2.7\),
complemented by cathode-strip chambers in the forward region, where the background is highest.
The muon trigger system covers the range \(|\eta| < 2.4\) with resistive-plate chambers in the barrel, and thin-gap chambers in the endcap regions.
 
Interesting events are selected by the first-level trigger system implemented in custom hardware,
followed by selections made by algorithms implemented in software in the high-level trigger~\cite{TRIG-2016-01}.
The first-level trigger accepts events from the \qty{40}{\MHz} bunch crossings at a rate below \qty{100}{\kHz},
which the high-level trigger further reduces in order to record events to disk at about \qty{1}{\kHz}.
 
An extensive software suite~\cite{ATL-SOFT-PUB-2021-001} is used in data simulation, in the reconstruction
and analysis of real and simulated data, in detector operations, and in the trigger and data acquisition
systems of the experiment.
 
\clearpage


\section{Signal models}
\label{sec:signals}

\begin{figure}[htbp]
\centering
\begin{subfigure}{0.27\textwidth}
\includegraphics[width=\textwidth]{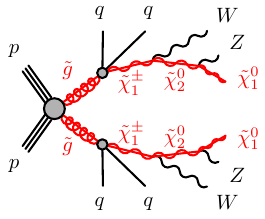}
\caption{}
\label{fig:models_GGwz}
\end{subfigure}
\begin{subfigure}{0.27\textwidth}
\includegraphics[width=\textwidth]{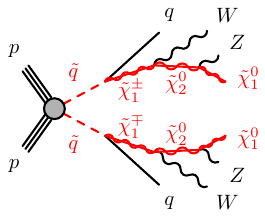}
\caption{}
\label{fig:models_SSwz}
\end{subfigure}
\begin{subfigure}{0.27\textwidth}
\includegraphics[width=\textwidth]{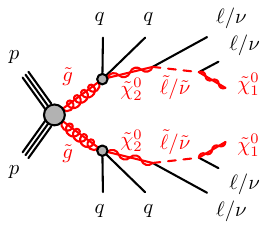}
\caption{}
\label{fig:models_GGslep}
\end{subfigure}
\begin{subfigure}{0.27\textwidth}
\includegraphics[width=\textwidth]{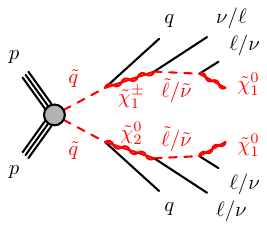}
\caption{}
\label{fig:models_SSslep}
\end{subfigure}
\begin{subfigure}{0.27\textwidth}
\includegraphics[width=\textwidth]{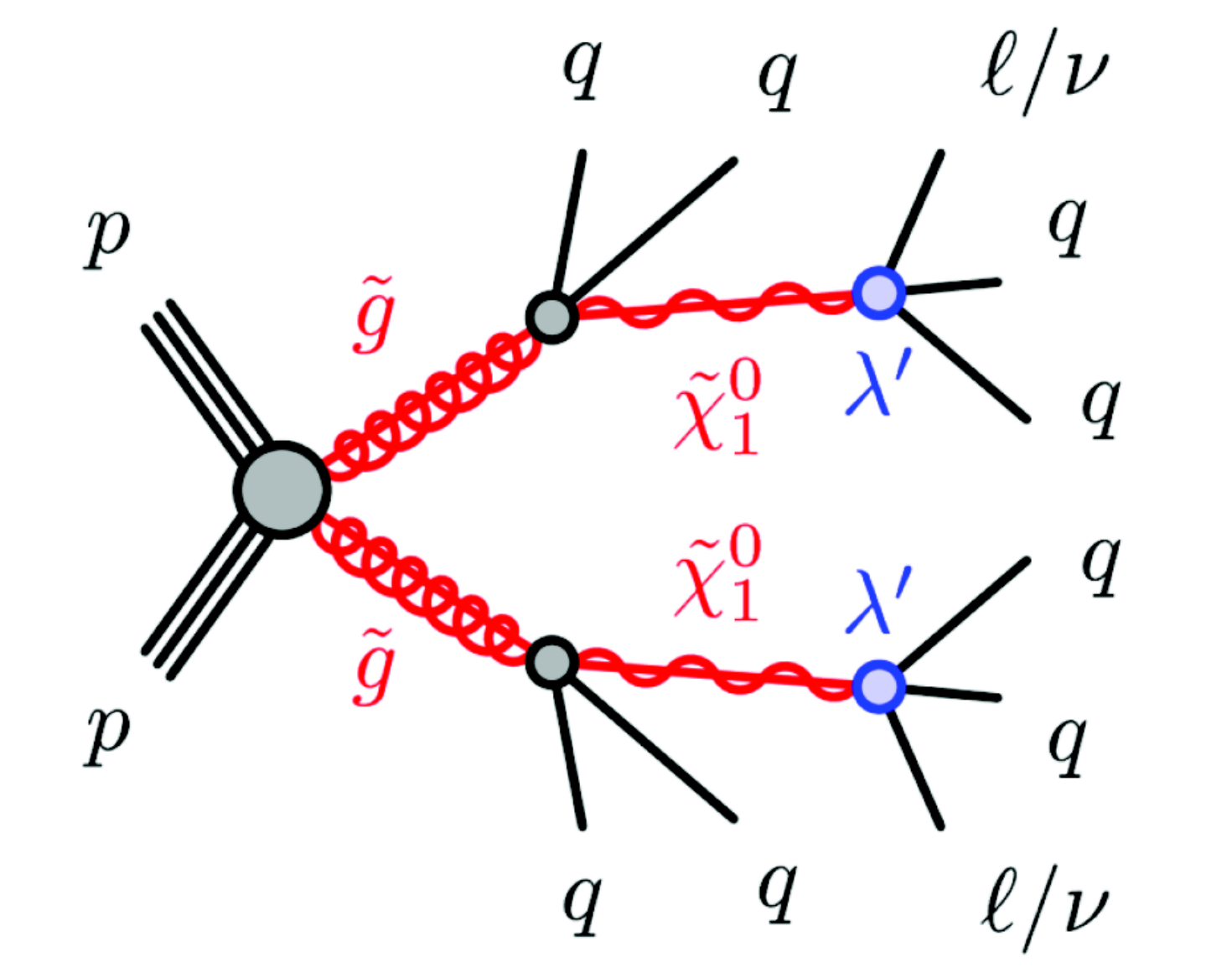}
\caption{}
\label{fig:models_GGrpv}
\end{subfigure}
\begin{subfigure}{0.27\textwidth}
\includegraphics[width=\textwidth]{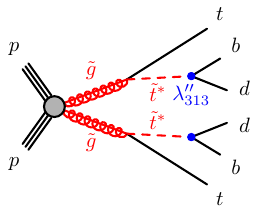}
\caption{}
\label{fig:RPV_UDD}
\end{subfigure}
\caption{Examples of sources of same-sign leptons which may arise in supersymmetric processes
and are targeted by the search regions of the analysis.
}
\label{fig:signals}
\end{figure}

This analysis considers experimental signatures arising
from the production of either gluino pairs
or squark--antisquark pairs $\tilde q\tilde q^{*}$ (with $q=u,d,s,c$).
Several gluino and squark decay modes are investigated,
and these are summarised in Figure~\ref{fig:signals}:
in most cases the first step of the decay chain
is to a non-stable neutralino or chargino \ggino
and SM quark(s) of first or second generation.
Various \ggino decay modes may lead to
final states featuring two same-sign leptons or three leptons.
Amongst those possibilities,
this analysis searches especially
for the following sources of electrons and muons:
\begin{itemize}
\item \ggino decays into SM gauge bosons and $\ninoone$ LSPs;
although the direct decays $\chinoonepm\to W^\pm\ninoone$
and $\ninotwo\to Z\ninoone$ are most efficiently probed
with other experimental signatures~\cite{SUSY-2018-22,SUSY-2018-10,SUSY-2018-05},
more favourable branching ratios to same-sign leptons are found
for cascade decays such as $\chinoonepm\to\ninotwo W^\pm\to \ninoone ZW^\pm$,
as illustrated in Figures~\ref{fig:models_GGwz} and~\ref{fig:models_SSwz},
which are thus the focus here.
\item \ggino decays into sleptons and subsequently into SM leptons and $\ninoone$ LSPs,
such as $\chinoonepm(\to \ell\snu/\slepton\nu)\to\ell\nu\ninoone$
or $\ninotwo(\to\slepton\ell)\to\ell^+\ell^-\ninoone$,
as illustrated in Figures~\ref{fig:models_GGslep} and~\ref{fig:models_SSslep}.
\item direct $\ninoone$ decay into SM leptons and quarks,
$\ninoone\to u\bar d\ell^+$, via a non-zero RPV coupling $\lambda'$
as illustrated in Figure~\ref{fig:models_GGrpv}.
\item $\gluino\to\stop\bar{t}\to \bar{t}bq$ decays
via a non-zero RPV coupling $\lambda''$,
shown in Figure~\ref{fig:RPV_UDD}.
\end{itemize}
 
The experimental sensitivity to such processes is assessed
in simplified models~\cite{Alwall:2008ve,Alwall:2008ag,Alves:2011wf}
where only the superpartners directly involved
in the process of interest are considered,
alternative production and decay modes are ignored,
and masses and mixings of superpartners are either varied freely
or fixed to chosen values.
 
For the cascade decays of charginos into pairs of SM bosons
(Figures~\ref{fig:models_GGwz} and~\ref{fig:models_SSwz}), the
gluino (or squark) and $\ninoone$ LSP masses are varied independently
to generate different scenarios.
The masses of intermediate superpartners are then set halfway in between,
i.e.\ $m(\chinoonepm)=(m(\gluino/\squark) + m(\ninoone))/2$
and
$m(\ninotwo)=(m(\chinoonepm) + m(\ninoone))/2$.
The decay of \chinoonepm and \ninotwo to leptons takes place
through real or virtual $W$/$Z$ bosons, depending on the mass difference from the LSP.
Quarks produced in the first step of the gluino decay chain
are assumed to be $u$, $d$, $s$ or $c$ with equal probability.
Due to the small Yukawa couplings for the first two quark generations,
the decays $\gluino\to qq\chinoonepm$ and $\squark\to q\chinoonepm$
are only relevant for a wino-like chargino~\cite{Martin:1997ns},
so only production of left-handed squarks is considered.
In minimal models with a neutralino LSP~\cite{Martin:1997ns},
such chargino cascade decays and a non-degenerate mass hierarchy
may occur when neutralinos are non-trivial admixtures,
but these would compete with several other, generally more favourable,
decay modes~\cite{Canepa:2020ntc}.
In gauge-mediated~\cite{Dine:1981gu,AlvarezGaume:1981wy,Nappi:1982hm}
SUSY breaking models, however,
this can be a more natural mode in the alternative form
$\chinoonepm\to W^\pm\ninoone\to W^\pm Z\gravino$
where the gravitino \gravino is the LSP.
Previous ATLAS searches using various signatures
~\cite{SUSY-2016-12,SUSY-2018-17,SUSY-2018-09,SUSY-2014-06}
probed gluino masses up to~\SI{2}{\TeV} and squark masses up to~\SI{630}{\GeV}
with these decay modes.
 
Similarly, for the decays of charginos and neutralinos into sleptons
(Figures~\ref{fig:models_GGslep} and~\ref{fig:models_SSslep}),
gluino (or squark) and $\ninoone$ LSP masses are varied independently,
while the masses of intermediate superpartners are set to
$m(\chinoonepm/\ninotwo)=(m(\gluino/\squark) + m(\ninoone))/2$
and
$m(\slepton/\snu)=(m(\chinoonepm/\ninotwo) + m(\ninoone))/2$.
Such scenarios arise in the mass spectra of models where sleptons are light,
and thus were searched for in early LHC analyses~\cite{
SUSY-2011-21,SUSY-2011-16,SUSY-2011-18}.
Gluinos are chosen to decay only into $\ninotwo$
while squarks are equally likely to decay into either $\chinoonepm$ or $\ninotwo$.
In both cases, the neutralino or chargino subsequently decays in an equiprobable way into
any of the six SM leptons, together with the
left-handed slepton appropriate for charge and lepton number conservation.
The latter decays exclusively into its SM partner and the LSP.
Previous ATLAS searches~\cite{
SUSY-2016-12,SUSY-2018-17,SUSY-2018-09,SUSY-2014-06}
probed gluino masses up to~\SI{2.2}{\TeV} and squark masses up to~\SI{850}{\GeV}.
The intermediate charginos or sleptons are also constrained
by previous analyses~\cite{SUSY-2018-32,SUSY-2018-04,CMS-SUS-17-009,CMS-SUS-21-001},
but for these particles the present analysis can probe masses
higher than those excluded by the direct-production searches,
thanks to their higher production rate in the cascade decays of gluinos or squarks.
 
Finally, event kinematics in the case of
neutralino decays via $\lambda^{'}$ RPV couplings
(Figure~\ref{fig:models_GGrpv})
are completely determined by the gluino and $\ninoone$ LSP masses,
which again are varied independently to generate different scenarios.
Without choosing explicit values,
it is assumed that the relevant $\lambda^{'}$ couplings
can be large enough to allow prompt \ninoone decay
into $\ell^- u\bar d$ or $\nu_{\ell}d\bar d$ ($\ell=e,\mu$),
or their charge conjugates,
with equal probabilities,
while evading low-energy experimental bounds~\cite{Barbier:2004ez}.
The \ninoone natural width also depends
on the mass of virtual squarks mediating the decay
and is set to~\SI{100}{\MeV}.
For such a scenario,
current experimental sensitivity~\cite{SUSY-2019-04}
excludes gluino masses up to~\SI{2.2}{\TeV}.
 
An additional scenario exemplified in Figure~\ref{fig:RPV_UDD} is considered,
in which pair-produced gluinos decay into the lightest top squark (and a SM top quark)
which itself decays via a suitable $\lambda''$ RPV coupling into a pair of quarks,
leading in half of the cases to a final state with two same-sign top quarks and up to four jets.
Such scenarios were highlighted in Ref.~\cite{Allanach:2012vj} especially,
and searches have been performed
in the $\bar{t}\bar{t}bbqq$~\cite{SUSY-2019-04, CMS-SUS-19-008,SUSY-2018-09}
and $\bar{t}\bar{t}qqqq$~\cite{SUSY-2016-14} final states.
In the present analysis, only the former case is addressed,
as it yields a clearer experimental signature;
for this scenario, Ref.~\cite{SUSY-2019-04} excluded gluino masses up to \SI{1.8}{\TeV}
for top squark masses around \SI{1}{\TeV}.
A set of benchmark models is generated by varying the gluino and top squark masses independently,
the latter being bounded from below
by existing constraints on $\stop\tilde{t}^{*}$ production~\cite{SUSY-2016-09}.
The top squarks are assumed to decay promptly.
A new set of search regions allows the sensitivity to be extended
beyond that reached in Ref.~\cite{SUSY-2018-09} with the same data.


\section{Data and samples of simulated events}
\label{sec:samples}

The results presented here are obtained
by analysing proton--proton collision data collected
during \RunTwo of the LHC at a centre-of-mass energy of~\SI{13}{\TeV}.
The number of simultaneous inelastic interactions
averages to 33.7 for the entire data set~\cite{DAPR-2018-01},
but exceeds 70 in a small fraction of the data.
Events recorded when parts of the detector were not functional
or reserved for detector commissioning or calibration purposes
are subsequently ignored,
leaving 95.6\% of the recorded data~\cite{DAPR-2018-01} available for analysis.
The integrated luminosity of this data set
amounts to \SI{139}{fb^{-1}}, with an uncertainty of $1.7\%$.
The latter is obtained~\cite{ATLAS-CONF-2019-021} using the LUCID-2 detector~\cite{LUCID2}
for the primary luminosity measurements.
 
Large samples of simulated events are also employed,
mainly to predict contributions from
SM processes with prompt\footnote{
Prompt leptons are defined are those produced
neither in the decay of a hadron,
nor radiatively in the fragmentation of quarks and gluons,
nor in the conversion of a photon
not originating from the electromagnetic shower of a charged lepton.
They may originate from the leptonic decay of a prompt $\tau$-lepton.
}
leptons to the search regions used in this analysis,
as well as those from hypothetical SUSY signal processes.
Other usages include validating assumptions employed in
data-based background estimation methods and,
more generally, assessing systematic uncertainties.
Those samples were obtained by simulating individual proton--proton collisions
for hard-interaction processes of interest
with the different combinations of MC event generators described below.
The events were then processed through a detailed simulation of the ATLAS detector~\cite{SOFT-2010-01}
based on \GEANT~\cite{Agostinelli:2002hh}.
In some cases, notably for BSM signal samples,
a faster simulation which relies on a parameterisation of
the calorimeter response~\cite{ATL-PHYS-PUB-2010-013} was used instead.
At this stage, additional minimum-bias interactions generated by \PYTHIA[8.186]
were simulated separately and overlaid on each simulated hard-interaction event to account for \pileup effects.
The response of the detector and its electronic readout chain is then emulated~\cite{SOFT-2010-01},
also accounting for effects from interactions in the previous and following bunch-crossings.
Reconstructed events are reweighted to reproduce the measured distributions
of the number of simultaneous interactions in different data-taking periods
and the measured effects of various sources of reconstruction inefficiency,
for example in the application of electron identification algorithms.
Specific kinematic variables, such as lepton momenta,
are smeared to reproduce the measured detector resolution.
 
The MC generators used to simulate the main SM processes of interest
are summarised in Table~\ref{tab:generators},
together with the selected parton shower algorithms,
the sets of tuned parameters (tunes),
and the sets of parton distribution functions (PDF).
When \PYTHIA was used, the decays of bottom and charm hadrons
were simulated with the \EVTGEN program~\cite{Lange:2001uf}.
Diboson processes~\cite{ATL-PHYS-PUB-2017-005} include all resonant and non-resonant
$pp\to 3\ell\nu/4\ell/\ell^{\pm}\ell^{\pm}\nu\nu$ processes of order $\alpha^4$
in the fine-structure constant, including Higgs boson contributions,
as well as the vector-boson scattering/fusion processes at order $\alpha^6$.
Triboson processes similarly include all the relevant resonant and non-resonant
processes with up to six charged leptons in the final state at order $\alpha^6$.
The associated production of \ttbar and an on-shell $W$ boson
includes a complementary sample generated at leading order (LO) in QCD
with the matrix elements of order $\alpha^3$.
Associated production of \ttbar and a pair of
same-flavour opposite-sign (SFOS) leptons
was generated for dilepton invariant masses as low as \SI{1}{\GeV}.
Other processes not identified individually in the table but included in the background estimates
comprise associated production of \ttbar and two vector or Higgs bosons,
associated production of single top quarks with one or two vector or Higgs bosons,
and production of three top quarks.
Fast detector simulation was employed for the
$4t$, $tH$, $tWH$, $\ttbar ZZ$, $\ttbar WH$, $\ttbar HH$ processes.

\begin{table}[t!]
\caption{
List of Monte Carlo event generators and their settings for the main  simulated samples
of SM processes.
When no reference is provided for the cross-section normalisation,
the one computed by the generator is used.
LO and NLO denote leading-order and next-to-leading-order calculations,
respectively; in some cases (indicated), matrix elements are used with different accuracies
depending on the number of additional parton emissions.
}
\label{tab:generators}
\scriptsize
\renewcommand{\arraystretch}{1.1}
\resizebox{\textwidth}{!}{
\begin{tabular}{l|l|l|l|l|l|l}
\hline\hline
Process & Generator & Computation & Parton shower & Cross-section & PDF set & Set of tuned \\
&           & order       &               & normalisation &         & parameters  \\
\hline
\hline
$t\bar{t}W$~\cite{ATL-PHYS-PUB-2020-024}
& \SHERPA[2.2.10]~\cite{Bothmann:2019yzt}
& NLO 0-1j + LO 2j
& \textsc{CSShower}~\cite{Schumann:2007mg}
& NLO
& \NNPDF[3.0nlo]~\cite{Ball:2014uwa}
& default \\
& + \OPENLOOPS~\cite{Buccioni:2019sur,Cascioli:2011va,Denner:2016kdg}
& + LO $\mathcal{O}(\alpha^3\alphas$)
& & & & \\
$t\bar{t}\ell^+\ell^-$~\cite{ATL-PHYS-PUB-2020-024}
& \SHERPA[2.2.1]~\cite{Bothmann:2019yzt}
& NLO
& \textsc{CSShower}~\cite{Gleisberg:2008fv,Schumann:2007mg}
& NLO
& \NNPDF[3.0nlo]~\cite{Ball:2014uwa}
& default \\
\quad $1<m_{\ell\ell}<\SI{5}{GeV}$
& MG5\_\AMCatNLO[2.3.3]~\cite{Alwall:2014hca}
& NLO
& \PYTHIA[8.212]~\cite{Sjostrand:2007gs}
& NLO
& \NNPDF[3.0nlo]~\cite{Ball:2014uwa}
& A14~\cite{ATL-PHYS-PUB-2014-021} \\
$t\bar{t}H$~\cite{ATL-PHYS-PUB-2016-005}
& \POWHEGBOX[v2]~\cite{Bothmann:2019yzt}
& NLO
& \PYTHIA[8.230]~\cite{Sjostrand:2007gs}
& NLO~\cite{deFlorian:2016spz}
& \NNPDF[3.0nlo]~\cite{Ball:2014uwa}
& A14~\cite{ATL-PHYS-PUB-2014-021} \\
$\ttbar\ttbar$~\cite{ATL-PHYS-PUB-2020-024}
& MG5\_\AMCatNLO[2.3.3]~\cite{Alwall:2014hca}
& NLO
& \PYTHIA[8.230]~\cite{Sjostrand:2007gs}
& NLO
& \NNPDF[3.1nlo]~\cite{Ball:2014uwa}
& A14~\cite{ATL-PHYS-PUB-2014-021} \\
& + \MADSPIN~\cite{Frixione:2007zp,Artoisenet:2012st}
& & & & & \\
Other $t/\ttbar+X$~\cite{ATL-PHYS-PUB-2020-024}
& MG5\_\AMCatNLO[2.3.3]~\cite{Alwall:2014hca}
& NLO or LO
& \PYTHIA[8.210-230]~\cite{Sjostrand:2007gs}
& NLO or LO
& \NNPDF[3.0/3.1nlo]~\cite{Ball:2014uwa}
& A14~\cite{ATL-PHYS-PUB-2014-021} \\
Diboson~\cite{ATL-PHYS-PUB-2017-005}
& \SHERPA[2.2.2]~\cite{Bothmann:2019yzt}
& NLO 0-1j + LO 2-3j
& \textsc{CSShower}~\cite{Gleisberg:2008fv,Schumann:2007mg}
& NLO
& \NNPDF[3.0nnlo]~\cite{Ball:2014uwa}
& default \\
& + \OPENLOOPS~\cite{Buccioni:2019sur,Cascioli:2011va,Denner:2016kdg}
& & & & & \\
Triboson~\cite{ATL-PHYS-PUB-2017-005}
& \SHERPA[2.2.1]~\cite{Bothmann:2019yzt}
& LO 0-1j
& \textsc{CSShower}~\cite{Gleisberg:2008fv,Schumann:2007mg}
& NLO
& \NNPDF[3.0nnlo]~\cite{Ball:2014uwa}
& default \\
\hline
$\ttbar$~\cite{ATL-PHYS-PUB-2016-020}
& \POWHEGBOX[v2]~\cite{Frixione:2007nw,Nason:2004rx,Frixione:2007vw,Alioli:2010xd}
& NLO
& \PYTHIA[8.230]~\cite{Sjostrand:2014zea}
& NNLO~\cite{ATL-PHYS-PUB-2016-020}
& \NNPDF[3.0nlo]~\cite{Ball:2014uwa}
& A14~\cite{ATL-PHYS-PUB-2014-021} \\
Single top ($s$-, $t$-channel)
& \POWHEGBOX[v2]~\cite{Re:2010bp,Nason:2004rx,Frixione:2007vw,Alioli:2010xd}
& NLO
& \PYTHIA[8.230]~\cite{Sjostrand:2014zea}
& NNLO~\cite{Campbell:2020fhf,PDF4LHCWorkingGroup:2022cjn}
& \NNPDF[3.0nlo]~\cite{Ball:2014uwa}
& A14~\cite{ATL-PHYS-PUB-2014-021} \\
\phantom{single top }($tW$)
& \SHERPA[2.2.7]~\cite{Bothmann:2019yzt}
& NLO
& \textsc{CSShower}~\cite{Gleisberg:2008fv,Schumann:2007mg}
& NNLO + NNLL~\cite{Kidonakis:2021vob}
& \NNPDF[3.0nnlo]~\cite{Ball:2014uwa}
& default \\
$W\to\ell\nu$, $Z/\gamma^*\to\ell\ell$~\cite{ATL-PHYS-PUB-2017-006}
& \SHERPA[2.2.11]~\cite{Bothmann:2019yzt}
& NLO 0-2j + LO 3-4j
& \textsc{CSShower}~\cite{Gleisberg:2008fv,Schumann:2007mg}
& NNLO~\cite{Anastasiou:2003ds,ATL-PHYS-PUB-2017-006}
& \NNPDF[3.0nnlo]~\cite{Ball:2014uwa}
& default \\
\hline\hline
\end{tabular}
}
\end{table}


The SUSY signal samples were generated with
MG5\_\AMCatNLO[2.6.2]~\cite{Alwall:2014hca} and the \NNPDF[2.3lo] PDF,
except for the $\gluino\to q\bar q WZ\ninoone$ and $\gluino\to t\bar{b}\bar{q}$ samples,
for which the \NNPDF[3.0lo] PDF was used.
The pair production of gluinos and squarks was simulated at LO,
complemented by matrix elements for up to two extra parton emissions.
Superpartners not involved in the model of interest
were decoupled by being assigned unreachable masses.
The decays of gluinos or squarks were factored out of the hard interaction
and simulated with~\PYTHIA[8.235]~\cite{Sjostrand:2014zea},
which was also used for the subsequent stages of the event generation
with the A14 tune~\cite{ATL-PHYS-PUB-2014-021}.
The matching between matrix elements and parton showers
followed the CKKW-L prescription~\cite{Lonnblad:2011xx},
with a matching scale set to one quarter of the gluino or squark mass.
For all models, the fast simulation was used to process the generated events.
Signal cross sections are calculated to approximate next-to-next-to-leading order
in the strong coupling constant, adding the resummation of soft gluon emission at
next-to-next-to-leading-logarithm accuracy
(approximate NNLO+NNLL)~\cite{
Beenakker:2016lwe,Beenakker:2014sma,Beenakker:2013mva,Beenakker:2011sf,
Beenakker:2009ha,Kulesza:2009kq,Kulesza:2008jb,Beenakker:1996ch}.
The nominal cross section and its uncertainty are derived using
the PDF4LHC15\_mc PDF set, following the recommendations of Ref.~\cite{Butterworth:2015oua}.


\section{Object selection}
\label{sec:objects}

Charged-particle tracks with $\abseta<2.5$ are reconstructed~\cite{
Cornelissen:2008,ATLAS-CONF-2012-042,Salzburger:2015sgq}
in the ID
and combined to form primary vertex candidates~\cite{
PERF-2015-01,ATL-PHYS-PUB-2015-026,ATL-PHYS-PUB-2019-015}.
The vertex with the largest $\sum\left(\pt^\text{track}\right)^2$ formed by at least two tracks
is taken to be the primary vertex and position of the hard-scattering interaction.
The transverse and longitudinal impact parameters of the tracks, denoted by
$d_0$ and $z_0$ respectively, are measured at their point of closest approach
to the beam line~\cite{IDTR-2019-05}.
Requirements placed on the ratio $|d_0^\text{sig}|$ of $d_0$ to its estimated
uncertainty $\sigma(d_0)$ and on the value of $|\zzsth|$ with respect to the
primary vertex are stated below.
 
Jets with $\abseta<4.5$ are reconstructed with
the \textsc{FastJet} implementation~\cite{Fastjet}
of the \antikt algorithm~\cite{Cacciari:2008gp},
with radius parameter $R=0.4$,
out of particle-flow objects~\cite{JETM-2018-05,PERF-2015-09}
combining calorimeter energy deposits~\cite{PERF-2014-07}
and ID tracks.
The jet's \pt, energy and mass are calibrated to the particle level~\cite{JETM-2018-05},
and only jets with $\pt>\SI{20}{\GeV}$ are retained.
Those originating from \pileup interactions,
according to a track-based discriminant~\cite{PERF-2014-03}, are rejected.
For selection criteria referring to the number of jets,
only jets with $\abseta<2.8$ are counted.
Around 0.5\% of the selected events are discarded due to the presence of
jets from sources~\cite{ATLAS-CONF-2015-029} other than $pp$ interactions.
Within the ID acceptance, jets containing bottom hadrons (referred to as \bjets)
are identified with the DL1r tagging algorithm~\cite{FTAG-2019-07},
which exploits the properties of reconstructed tracks and secondary vertices.
The analysis selects true \bjets with an estimated 70\% efficiency in \ttbar events
while rejecting 99.8\% of other jets free of charm hadrons or $\tau$-leptons~\cite{FTAG-2019-07}.
 
Muons with $|\eta|<2.5$ and $\pt>\SI{10}{\GeV}$ are used.
They are obtained~\cite{MUON-2018-03} from an iterated track fit of ID and MS hits.
Momentum corrections~\cite{MUON-2022-01} compensate for detector misalignments.
The \enquote{Medium} quality criteria defined in Ref.~\cite{MUON-2018-03} are applied,
and \pileup muons are rejected by requiring $|\zzsth|<\SI{0.5}{mm}$.
The candidates satisfying these requirements are referred to as baseline muons.
Around 0.1\% of events contain a muon from a cosmic-ray shower
or with poor expected momentum resolution, and are rejected.
Prompt muons are further distinguished from background sources
with a requirement $|d_0^\text{sig}|<3$
and an isolation criterion~\cite{MUON-2018-03}
consisting of an upper bound of 6\%, relative to the muon \pt,
on the summed $|\pt|$ of suitable ID tracks\footnote{
These have $\pt>\SI{1}{\GeV}$, originate from the primary vertex,
and exclude the muon's own track; more details can be found in Ref.~\cite{MUON-2018-03}.
}
within $\Delta R<\operatorname{min}(0.3, \SI{10}{\GeV}/\pt(\mu))$.
They are referred to as signal muons.
 
Electrons with $|\eta|<2.47$ and $\pt>\SI{10}{\GeV}$ are used.
They are reconstructed~\cite{EGAM-2018-01}
from clustered energy deposits in the electromagnetic calorimeter
matched to an ID track re-fitted to account for bremsstrahlung losses.
Their momentum is determined by a calibration procedure~\cite{PERF-2017-03,EGAM-2018-01}
based on boosted decision trees (BDTs).
Electrons within the calorimeter's barrel-to-endcap transition region $1.37<|\eta|<1.52$ are not considered.
Electrons from most background sources are rejected
with a likelihood discriminant~\cite{EGAM-2018-01}
built from information about the development of
the electron shower in the calorimeter,
its compatibility with the matched track,
and particle identification in the TRT detector.
Electrons with $|\zzsth|>\SI{0.5}{mm}$ are rejected.
The candidates satisfying these requirements are referred to as baseline electrons.
Non-prompt or fake electrons are further suppressed
by keeping only electrons within $|\eta|<2.0$,
imposing the tighter \enquote{MediumLH} identification criteria~\cite{EGAM-2018-01},
a combination of track-based and calorimeter-based isolation criteria~\cite{EGAM-2018-01},
and requiring $|d_0^\text{sig}|<5$.
The first isolation criterion is similar to the one used for muons,
while the other consists of an upper bound of 6\%, relative to the electron \pt,
on the sum of other calorimeter energy deposits within $\Delta{R}<0.2$.
Electrons very likely to have a wrongly assigned charge are identified and then rejected
using the ECIDS discriminant~\cite{EGAM-2018-01},
a BDT which is based on the properties of the electron track
and which accepts $98\%$ of simulated $Z\to ee$ decay electrons
while rejecting $90\%$ of those with the wrong charge.
The remaining candidates are referred to as signal electrons.
 
An overlap removal procedure is applied to baseline lepton candidates and jets
to avoid treating the same detector signals as multiple objects.
Jets close to electrons ($\Delta{R}<0.2$) are removed,
unless they are classified as \bjets with $\pt<\SI{100}{\GeV}$, and
so are jets close to muons ($\Delta{R}<0.2$ or sharing an ID track)
if they have less than three associated ID tracks.
Leptons close to remaining jets
($\Delta{R}<\operatorname{min}(0.4,0.1+\SI{9.6}{\GeV}/\pt(\ell))$)
are removed, as they are likely to be fake or non-prompt;
the $\Delta{R}$ bound is smaller at high \pt
to retain leptons from boosted top quark or \ninoone decays.
Electrons close to muons ($\Delta{R}<0.01$)
or to higher-\pt electrons ($\Delta{R}<0.05$) are removed.
Throughout this procedure, $\Delta{R}$ is calculated
with the rapidity $y$ rather than the pseudorapidity $\eta$.
 
The missing transverse momentum \ptmiss\
and its magnitude \met are reconstructed~\cite{ATLAS-CONF-2018-023}
from selected electrons, muons and jets prior to overlap removal,
together with reconstructed photons ($\pt>\SI{25}{\GeV}$, $|\eta|<2.37$)
satisfying \enquote{Tight} identification criteria~\cite{EGAM-2018-01}
and a track-based \enquote{soft term} consisting of softer contributions associated with
the primary vertex but not included in aforementioned objects.
The \met reconstruction employs its own overlap removal procedure.


\section{Event selection}
\label{sec:strategy}


Events with $\met<\SI{250}{\GeV}$ were selected using  dilepton triggers~\cite{TRIG-2018-05,TRIG-2018-01}.
Since the luminosity increased during \RunTwo, the lepton \pt\ thresholds were raised in steps to a maximum of \SI{24}{\GeV} for triggers requiring two electrons,  $\SI{22}{\GeV}$ for triggers requiring two muons, and $\SI{17}{\GeV}$ (\SI{14}{\GeV}) for the electron (muon) in different-flavour dilepton triggers.
For events with $\met>\SI{250}{\GeV}$, a logical OR of these dilepton triggers and $\met$ triggers~\cite{TRIG-2019-01} was used.  Events are preselected by requiring exactly two signal leptons with the same electric charge or at least three  signal leptons\footnote{In the following, unless otherwise stated, \enquote{leptons} refers to signal leptons.} without any charge requirement.

Multiple signal regions (SRs) were defined with the goal of maximising the sensitivity to the signal models shown in Figure~\ref{fig:signals}.  These SRs are not exclusive and can overlap.
They are primarily built on requirements placed on the number of signal ($\NSigLep$) and/or baseline ($\NBaseLep$) leptons and their relative charges,  the number of $b$-jets ($\nb$) with $p_\text{T}>20$~\GeV,  the number of jets ($\nJ$) with $p_\text{T}$ above 25, 40 or 50~\GeV, regardless of their flavour, and also apply independent selections to a series of observables sensitive to kinematic differences between signal and background events.
For each SR, numerous candidate observables were assessed, but only those bringing the best sequential improvement
in expected sensitivity were retained; this explains the variability in the chosen selection variables for the different SRs.
One of the most sensitive observables used across all SRs is the effective mass, $\meff$, which aids in establishing the mass scale of the processes being probed, and is defined as the scalar sum of the $p_\text{T}$ of the jets and leptons and the $\met$ of the event,
 
\begin{equation}
\label{eq:meff}
\meff = \HTJets+ \HTLep+  \met.
\end{equation}

\subsection{RPC SRs}
 
SRs targeting the RPC models  are shown in Tables~\ref{tab:final_GGWZSRs}--\ref{tab:final_rpcSSslep}.
Depending on the model, RPC SRs can  require  events with at least  two leptons with the same electric charge, or  three or more leptons.   SRs with a three-lepton selection accept any charge combination for the three leptons.
A veto on $b$-tagged jets is imposed in order to reduce SM backgrounds with top quarks.
For those signal models not involving the presence of a  $Z$ boson in the final state (Figures~\ref{fig:models_GGslep} and~\ref{fig:models_SSslep}), events in which the invariant mass of any pair of two same-flavour opposite-sign leptons is compatible with the $Z$ boson mass ($\mSFOS \in[81,101]$~\GeV) are vetoed.

SRs targeting the benchmark models where pair production of gluinos (Figure~\ref{fig:models_GGwz}) and squarks (Figure~\ref{fig:models_SSwz}) leads to cascade decays of charginos into pairs of SM bosons are named SRGGWZ  and SRSSWZ, respectively.
Multiple SRs are defined for each benchmark model, specifically tailored to target the different  neutralino  and gluino (or squark) mass-splitting scenarios.
Those different scenarios are identified by a suffix \enquote{-L}, \enquote{-M}, or \enquote{-H}  --standing for low, medium and high mass splitting-- in the SR name, with \enquote{H} being the scenario where $m_{\tilde{g}(\tilde{q})}\gg m_{\tilde{\chi}_1^0}$, \enquote{L}  denoting the scenario where  $m_{\tilde{g}(\tilde{q})}\approx m_{\tilde{\chi}_1^0}$, and \enquote{M} referring to the SR defined for the intermediate phase space between \enquote{L} and \enquote{H}.   For those benchmark models where squarks are pair produced, the SR \enquote{-M} is further split into a region in the intermediate phase space close to the low mass splitting, \enquote{-ML},  and another one close to the high mass splitting,  \enquote{-MH}.  Requirements are placed on the  jet multiplicity, $\met$, $\meff$ or   $\HTJets$, depending on the kinematics of the objects generated in the different mass-splitting scenarios.
Additional selection criteria are applied to other observables that were found to discriminate strongly between signal and background events.
Those variables are the ratios of the different terms present in Eq.~(\ref{eq:meff}), i.e.\ \MetOverHTJets,  \MetOverMeff, \MetOverHTLep,  \MeffOverHTLep,  \HTLepOverHTJets;  the azimuthal separation between the  system formed by the two leading leptons and the direction of the missing transverse momentum of the event, $\DPhillMet$;    and the $\met$ significance,~\metsig~\cite{ATLAS-CONF-2018-038}.

\begin{table}[ht!]
\centering
\caption{Definition of the  signal regions used for the RPC model  shown in Figure~\ref{fig:models_GGwz}, where pair production of  gluinos leads to cascade decays of charginos into pairs of SM bosons.
}
\label{tab:final_GGWZSRs}
\small
\setlength{\tabcolsep}{3pt}
\def\arraystretch{1.2}
\begin{tabular}{l|c|c|c|c|c|c|c|c}
\hline
\hline
SR   name       & \NSigLep (\NBaseLep) & \nb & $\nJ$ & $\pTjet$ [$\GeV$]  & \met[$\GeV$]    & \meff[$\GeV$]   & $\DPhillMet$ & \metsig \\
\hline\hline
SRGGWZ-L    & $\ge 2$ ($\ge 3$)           &      \multirow{3}{*}{$=0$}        & $\ge6$   & $>25$           & $>200$  & $>8\times\HTLep$ & $>0.2$       & $>6$ \\
 
\cline{1-2}
\cline{4-9}
SRGGWZ-M    & $\ge 2$ (--)           &          & $\ge6$   & $>40$           & $>190$  & $>1300$  & $>0.8$       & --  \\
\cline{1-2}
\cline{4-9}
SRGGWZ-H   & $\ge 2$ (--)           &    & $\ge6$   & $>40$           & $>150$  & $>2100$ & --        & --\\
\hline
\hline
\end{tabular}
\end{table}
 
\begin{table}[h!]
\centering
\caption{
Definition of the  signal regions used for the RPC model  shown in Figure~\ref{fig:models_SSwz}, where pair production of  squarks leads to cascade decays of charginos into pairs of SM bosons.
}
\small
\setlength{\tabcolsep}{3pt}
\def\arraystretch{1.2}
 
\begin{tabular}{l|c|c|c|c|c|c|c|c|c}
\hline\hline
SR  name & \NSigLep   & $\nb$ & $\nJ$ &  $\pTjet$  [$\GeV$]  & \met[$\GeV$] & \meff[$\GeV$] & \MetOverHTLep & \HTLepOverHTJets & $n_{Z\to\ell^+\ell^-}$\\
\hline
\hline
SRSSWZ-L &   \multirow{4}{*}{ $\geq 3$}   &  \multirow{4}{*}{$=0$} & $\geq 4$ & $>25$  & $>0.2\times\meff$        & --    & --      & $<0.2$     & $=0^\dagger$ \\
 
\cline{1-1}
\cline{4-10}
 
SRSSWZ-ML &     &   & $\geq 6$ & $>25$  & $>150$  & $>800$  & $>1.2$ & $<0.3$  & $\geq 1^\dagger$\\
 
\cline{1-1}
\cline{4-10}
SRSSWZ-MH &     &  & $\geq 5$ & $>40$  & $>200$  & $>900$ & $>1.1$ & $<0.4$  & $\geq 1^\dagger$ \\
 
\cline{1-1}
\cline{4-10}
SRSSWZ-H  &     &  & $\geq 5$ & $>40$ & $>250$  & $>1500$\phantom{0} & $>0.3$        & $<0.7$  & -- \\
\hline\hline
\multicolumn{9}{l}{$^\dagger$: based on number of SFOS pairs with $81 < \mSFOS <\SI{101}{\GeV}$}\\
\hline\hline
\end{tabular}
\end{table}

\FloatBarrier

SRs targeting the  benchmark models where pair production of gluinos (Figure~\ref{fig:models_GGslep}) and squarks (Figure~\ref{fig:models_SSslep}) leads to decays of charginos and neutralinos into sleptons are named SRGGSlep  and SRSSSlep, respectively. Multiple SRs are defined for the different mass-splitting scenarios and named as per the convention described previously.  An extra SR  (SRSSSlep-H (loose)) is defined for the RPC model shown in Figure~\ref{fig:models_SSslep}, using the same selection criteria as for  SRSSSlep-H, but with the  $\meff$ requirement relaxed to 1~\TeV\ to allow for a binned fit in the model-dependent interpretation (see Section~\ref{sec:statistical}).  SRs are defined in terms of the variables previously described, with requirements also placed  on the $p_\text{T}$ of the leading and sub-leading leptons, denoted by  \ptLeadL and \ptSubLeadL respectively,  and on the angular distance between the two leading leptons, \DRLL.

\begin{table}[ht!]
\centering
\caption{Definition of the  signal regions used for the RPC model shown in Figure~\ref{fig:models_GGslep}, where pair production of gluinos leads to decays of charginos and neutralinos into sleptons.
}
\small
\setlength{\tabcolsep}{3pt}
\def\arraystretch{1.2}
\begin{tabular}{l|c|c|c|c|c|c|c|c}
\hline
\hline
SR          name  & \NSigLep    & \nb & $\nJ$ & \pTjet[$\GeV$] & \met[$\GeV$] &  \MetOverHTJets\ & \ptSubLeadL[$\GeV$]  & Other\\
\hline\hline
SRGGSlep-L & \multirow{3}{*}{  $\ge 3^\dagger$}  &  \multirow{3}{*}{$=0$} & \multirow{3}{*}{ $\ge 4$} &  \multirow{3}{*}{$\ge 40$} & -- & $>0.4$ & $>30$ & $\MetOverHTLep>1.4$  \\
\cline{6-9}
SRGGSlep-M & & & & & $>150$ & $>0.3$ & $>70$ & $\DPhillMet>0.7$\\
\cline{6-9}
SRGGSlep-H & & & & & $>100$ & -- & -- & $\HTJets>\SI{1200}{\GeV}$ \\
\hline
\hline
\multicolumn{9}{l}{$^\dagger$: SFOS pairs with $81 < \mSFOS <\SI{101}{\GeV}$
are not allowed}\\
\hline
\hline
\end{tabular}
\end{table}

\begin{table}[h!]
\centering
\caption{
Definition of the  signal regions used for the RPC model shown in Figure~\ref{fig:models_SSslep},
where pair production of squarks leads to decays of charginos and neutralinos into sleptons.
Requirements on $\pt^\ell$ apply to all three leptons.
}
\label{tab:final_rpcSSslep}
\small
\setlength{\tabcolsep}{3pt}
\def\arraystretch{1.2}
\begin{tabular}{l|c|c|c|c|c|c|c|c}
\hline\hline
SR name  & \NSigLep & $\pt^\ell$ [$\GeV$]   & $\nb$ & $\nJ$ &  $\pTjet$  [$\GeV$]   & \met[$\GeV$]  & \meff[$\GeV$] & \DPhillMet  \\
\cline{2-9}
& \multicolumn{8}{c}{other requirements}\\
\hline
\hline
\multirow{2}{*}{SRSSSlep-L}
&  $=3^{*}$    & $<60$ & $=0$ & $\geq 3$ & $>60, 60, 25$& $>100$   & \phantom{0}$>600$ & $>1.4$\\
\cline{2-9}
& \multicolumn{8}{c}{$\HTLepOverHTJets<0.6$}\\
\hline
\multirow{2}{*}{SRSSSlep-ML}
&  $=3^{*}$ & $> 30$ & $=0$ & $\geq 3$ & $>60,60,25$ & $>100$  & \phantom{0}$>700$ & $>1.4$\\
\cline{2-9}
& \multicolumn{8}{c}{$\MetOverHTLep>0.7$,\quad$\HTLepOverHTJets<0.6$}\\
\hline
\multirow{2}{*}{SRSSSlep-MH}
& $=3^{*}$ & $>40$ & $=0$ & $\geq 2$ & $>60$ & $>200$  & $>1000$ & $>0.5$\\
\cline{2-9}
& \multicolumn{8}{c}{$\MetOverHTLep>0.7$,\quad$\DRLL>0.2$}\\
\hline
\multirow{2}{*}{SRSSSlep-H}
& $=3^{*}$ & $>40$ & $=0$ & $\geq 2$ & $>60$ & $>200$  &  $>2000$  & $>0.3$\\
\cline{2-9}
& \multicolumn{8}{c}{$\DRLL>0.5$}\\
\hline
\multirow{2}{*}{SRSSSlep-H (loose)}
& $=3^*$ & $>40$ & $=0$ & $\geq 2$ & $>60$ & $>200$   & $>1000$ & $>0.3$\\
\cline{2-9}
& \multicolumn{8}{c}{$\DRLL>0.5$}\\
\hline\hline
\multicolumn{9}{l}{$^*$:
additional baseline leptons are not allowed,
nor SFOS pairs with $81 < \mSFOS <\SI{101}{\GeV}$}\\
\hline\hline
\end{tabular}
\end{table}

\FloatBarrier
 
\subsection{RPV SRs}

SRs targeting the  RPV models are shown in Tables \ref{tab:final_rpvLQDSRs} and \ref{tab:final_rpvUDDSRs}.
The SRs are named SRLQD  and SRUDD, where the former corresponds to the SR defined for the model where the neutralino decays via the $\lambda^\prime$ RPV coupling of LQD type (Figure~\ref{fig:models_GGrpv}), and the latter to the SRs defined for the model where gluinos decay via top squarks and the $\lambda^{\prime\prime}$ RPV coupling of UDD type (Figure~\ref{fig:RPV_UDD}).
For the latter, the suffix appended to the SR name indicates the requirement on the number of $b$-jets.
Those SRs require exactly two leptons with the same electric charge,  high jet multiplicity,  and high  $\HTJets$  or high $\meff$.     No requirement is placed on the $\met$ as no neutralinos are expected in the final state of  RPV model events.

\begin{table}[h!]
\centering
\caption{
Definition of the  signal region used for the RPV model  shown in  Figure~\ref{fig:models_GGrpv}, where the neutralino decays via the $\lambda^\prime$ RPV coupling of LQD type.
}
\label{tab:final_rpvLQDSRs}
\small
\setlength{\tabcolsep}{3pt}
\def\arraystretch{1.2}
\begin{tabular}{l|c|c|c|c|c}
\hline\hline
SR     name                             & \NSigLep  & $\nb$  & $\nJ$      &  $\pTjet$  [$\GeV$]       & \meff[$\GeV$]  \\
\hline
\hline
SRLQD&  $=2$    & $-$     & $\geq 5$ & $>50$                  & $>2600$ \\
\hline\hline
\end{tabular}
\end{table}

\begin{table}[h!]
\centering
\caption{
Definition of the  signal regions used for the RPV model shown in Figure~\ref{fig:RPV_UDD}, where gluinos decay via top squarks and the $\lambda^{\prime\prime}$ RPV coupling of UDD type.
}
\label{tab:final_rpvUDDSRs}
\small
\setlength{\tabcolsep}{3pt}
\def\arraystretch{1.2}
\begin{tabular}{l|c|c|c|c|c|c}
\hline\hline
SR           name            &  \NSigLep & $\nb$    &  $\nJ$       &  $\pTjet$  [$\GeV$]   & \meff[$\GeV$] & \HTJets\  [$\GeV$] \\
\hline
\hline
SRUDD-1b   & \multirow{4}{*}{$=2$}     &  $=1$  &  $\geq$ 6  & $>50$               & $-$               & $>1600$  \\
\cline{1-1}
\cline{3-7}
SRUDD-2b&     &  $=2$   &  $\geq$ 2  & $>25$               & $-$               & $>1700$   \\
\cline{1-1}
\cline{3-7}
SRUDD-ge2b &   & $\geq$ 2 & $\geq$ 5  & $>50$               & $-$               & $>1600$   \\
\cline{1-1}
\cline{3-7}
SRUDD-ge3b  &    & $\geq$ 3  & $\geq$ 4  & $>50$             & $>1600$ & $-$         \\
\hline\hline
\end{tabular}
\end{table}


\section{Background estimation}
\label{sec:background}

Background contributions to the search regions
can be divided into three categories,
consisting in SM processes with genuine same-sign leptons in the final state,
and other SM processes forming same-sign lepton pairs because of
the incorrect reconstruction of an electron's charge
or the presence of fake or non-prompt (F/NP) leptons.
The following three sub-sections detail the estimation method for each category:
processes with genuine same-sign leptons are estimated with MC simulations,
aided by a single control region in the data,
while for the other two,
data events with specific lepton selection criteria are exploited.
Comparisons between background predictions and observed data in
carefully designed validation regions
are presented in a fourth sub-section.
 
\subsection{SM processes with prompt same-sign leptons}
 
The largest contribution to the RPC SRs
originates from $WZ$\,+\,jets with both bosons decaying into leptons.
All these SRs veto events with $b$-jets,
which suppresses processes involving top quarks.
For $WZ$\,+\,jets, a control region intermediate in jet multiplicity is employed,
referred to as CRWZ2j.
It is defined with three signal leptons (and no fourth baseline lepton),
two of them forming a SFOS pair with invariant mass in the range $81<\mSFOS<\SI{101}{GeV}$.
The two same-sign leptons of the triplet must have $\pt>\SI{15}{GeV}$,
and the sum of the three leptons' $\pt$ must exceed \SI{130}{GeV}.
Events must contain either two or three jets with $\pt>\SI{25}{GeV}$, with no \bjet present.
Contributions from hypothetical BSM processes
are reduced by requiring $30<\met<\SI{150}{GeV}$ and $\meff<\SI{1.5}{TeV}$.
This selection results in
an estimated purity exceeding 85\%,
with the remainder dominated by other multiboson processes.
The overall normalisation of the $WZ$\,+\,jets MC prediction is then
treated as a free parameter and set by fitting
the observed data yields in CRWZ2j.
This is realised with a maximum likelihood fit
formed by the joint probability of expected background contributions to CRWZ2j with their uncertainties.
In the case of signal-strength hypothesis testing,
the joint probability is expanded to include contributions to the signal regions of interest,
as well as the tested signal contribution to all regions.
The procedure is fully detailed in Section~\ref{sec:statistical}.
It is found that a scale factor of $0.86\pm 0.05$
relative to the MC prediction using the theoretical inclusive NLO cross section
best accommodates the observed data in CRWZ2j.
This below-unity value is consistent with past observations~\cite{STDM-2017-23,STDM-2018-03}.
In the RPV SRs, no simultaneous CR-SR fit is performed
and the $WZ$\,+\,jets normalisation is determined by CRWZ2j alone.
This sequential approach does not propagate correlations
among the $WZ$\,+\,jets uncertainties for RPV SRs,
but this does not affect the results because the $WZ$\,+\,jets background is subdominant in those SRs.
 
Other SM background processes with prompt same-sign leptons
are estimated by normalising the simulated samples to the theoretical cross section.
Some weaker sources of same-sign leptons are ignored.
Those include radiative top quark decays $t\to b\ell^+\nu\ell^+\ell^-$ in $\ttbar$ events,
found to be negligible for the analyses in Refs.~\cite{HIGG-2018-13,SUSY-2018-09},
and multiple parton scattering, for which similar conclusions were reached in Ref.~\cite{SUSY-2018-09}.

\subsection{Electrons with incorrect charge}
 
Electrons and positrons are distinguished by the curvature of their ID tracks,
which may be error-prone if they start radiating early in the ID.
The probability $\pi$ for signal electrons to exhibit a charge flip
is around $0.06\%$ in the validation region with the largest charge-flip contribution.
It varies by several orders of magnitude with $|\eta|$ and \pt
as illustrated in Ref.~\cite{EGAM-2018-01}.
Contributions of charge-flip background to the various analysis regions
are estimated by selecting data events with opposite-sign leptons
and weighting them according to the known $\pi(|\eta|,\pt)$ values.
The latter are measured in simulated $\ttbar$ events,
and multiplied by factors $\gamma(|\eta|,\pt)$ correcting the known mismodelling.
The factors $\gamma$, assumed to be process-independent, are determined~\cite{EGAM-2018-01}
from the rates of opposite- and same-sign dielectron pairs observed in $Z\to ee$ decays in data and MC events.
They are found to be within 20\% of unity.
 
The dominant uncertainties in the predicted charge-flip yields
are those arising from the measurement of the $\gamma$ corrections,
which is statistically limited and affected by a significant background.
The predicted yields suffer from a typical uncertainty of 50\%,
which, however, does not impact the results of the present analysis
since they represent at most 7\% of the total background in any of the SRs.

\subsection{Fake and non-prompt leptons}
 
Non-prompt leptons arising from hadron decays or photon conversions,
as well as hadrons misreconstructed as electrons,
may survive the identification criteria in Section~\ref{sec:objects} at a low rate.
When combined with a prompt lepton in an event, they may form
a same-sign lepton pair.
This source of background is estimated primarily with the matrix method
summarised in the following paragraphs,
and its prediction is then compared
with that of a corrected simulation~\cite{SUSY-2016-14,ATLAS-CONF-2022-057}.
 
The matrix method~\cite{D0:1999qdf}
exploits the different efficiency of the identification and isolation requirements
when applied to F/NP leptons.
In a given region of interest, data events are selected by applying
lepton selection criteria that are looser than those defining signal leptons,
and then categorising those events according to the number of actual signal leptons.
A fully determined system of linear equations can then be constructed,
relating the numbers of such categorised events to the unknown numbers of events
with only prompt leptons, or exactly one F/NP lepton, etc.,
where the coefficients are functions of the probabilities $\varepsilon$ and $\zeta$
for loose prompt leptons or F/NP leptons to also satisfy the nominal criteria.
This can be illustrated for events with a single lepton by the matrix equation:
 
\begin{equation*}
\begin{pmatrix} n_\text{signal} \\ n_\text{all} \end{pmatrix}
=
\begin{pmatrix} 1 & 1 \\ \varepsilon^{-1} & \zeta^{-1} \end{pmatrix}
\cdot
\begin{pmatrix} n_\text{signal, prompt} \\ n_\text{signal, F/NP} \end{pmatrix}.
\end{equation*}
 
The desired estimate $n_\text{signal, F/NP}$ can then be obtained easily.
This approach can be readily extended to final states with multiple leptons.
The general method is described in detail in Ref.~\cite{EGAM-2019-01}, and
the present analysis relies upon the software implementation referenced therein.
The sample of loosely selected leptons comprises
the subset of baseline leptons after overlap removal,
which for muons satisfy $|d_0^\text{sig}|<7$,
and for electrons satisfy the ECIDS requirement in Section~\ref{sec:objects}
and are restricted to $|\eta|<2.0$.
The estimated contribution of charge-flip electrons is subtracted from all terms.
 
The probabilities $\varepsilon(|\eta|,\pt)$ are calculated
with simulated $\ttbar$ decays to leptons.
The probabilities $\zeta(\pt)$ are measured in data~\cite{SUSY-2018-09}
within the range $10<\pt<\SI{75}{GeV}$ in regions enriched in \ttbar events
with one or two prompt leptons and one F/NP lepton forming a same-sign pair,
which is also the leading contributor of F/NP leptons to the SRs.
Separate measurements are made depending on the number of \bjets in the event ($\leq 1$ or $\ge 2$)
and whether the electron satisfies the criteria~\cite{TRIG-2018-05} to trigger the event recording.
The undesirable contributions of $WZ$\,+\,jets and $t\bar{t}W$ to these measurement regions, estimated with MC samples,
are scaled by correction factors of 0.84 and 1.19 respectively,
based on observed data in CRWZ2j and the region VRTTV defined in Table~\ref{tab:VRs_defs}.
For muons, $\varepsilon$ is found to increase with \pt
from 75\% at \SI{10}{GeV} to 98\% at \SI{70}{GeV},
while $\zeta$ varies between 10\% and 15\% in the most relevant range, $10<\pt<\SI{40}{GeV}$,
reaching its lowest values at around \SI{20}{GeV}.
For electrons, $\varepsilon$ also increases with \pt, from 50\% at \SI{10}{GeV} to 97\% at \SI{100}{GeV},
while $\zeta$ varies between 5\% and 10\% in the aforementioned \pt range.
When the electron satisfies the trigger criteria, $\zeta$ values can be up to twice as large.
Over the whole measurement range and for both lepton flavours,
$\varepsilon$ is always found to be much larger than $\zeta$,
a necessary condition for the applicability of the matrix method~\cite{EGAM-2019-01}.
In the SRGGWZ-L region, which only selects events containing
a third baseline lepton in addition to the pair of same-sign signal leptons,
the estimate is calculated with the reasonable assumption
that the F/NP lepton is part of the latter pair.
 
The same conservative systematic uncertainties as in Ref.~\cite{SUSY-2018-09} are used
to account for contamination from prompt same-sign lepton processes
in the measurement regions and for the assumption that
$\varepsilon$ and $\zeta$ can be used outside of the regions in which they are measured.
The latter is assessed primarily with $t\bar{t}$ simulation,
such that the uncertainties cover the effects of having unknown relative contributions
of different sources of F/NP leptons
and residual variations of the probabilities as function of the environment
that are not captured by the parameterisation as function of $\pt$.
This procedure leads to uncertainties in $\zeta$ ranging from $30\%$ at
lower \pt for both flavours, up to 80\% for high-\pt muons.
Combined with statistical uncertainties in the $\zeta$ measurements,
these translate for example into systematic uncertainties in the predicted F/NP yields
of 44\% and 75\% in the SRSSWZ-L and SRUDD-ge2b signal regions, respectively,
which contain a significant F/NP contribution.
They are in most cases comparable in size to
the corresponding statistical uncertainties in the predicted F/NP SR yields.
 
The SR yields predicted with the matrix method were cross-checked
against those predicted by MC simulations of $\ttbar$ and $W/Z$\,+\,jets processes.
For the latter, simulated events with F/NP leptons are weighted
by three correction factors extracted from fits to data in appropriate regions:
one for F/NP electrons, one for non-prompt muons from bottom hadron decays,
and one for other sources of non-prompt muons.
The estimates are compatible within uncertainties with the ones obtained with the matrix method.
 
\subsection{Validation of the background estimates}
 
\begin{table}[t!]
\centering
\caption{
The definitions of validation regions used to check the
accuracy of the SM background predictions.
Requirements are placed on signal leptons, jets,
and some of the event-level variables defined in Section~\ref{sec:strategy}.
}
\label{tab:VRs_defs}
\small
\setlength{\tabcolsep}{5pt}
\renewcommand{\arraystretch}{1.1}
\begin{tabular}{l|c|c|c|c|c|c}
\hline\hline
& \NSigLep & \nb & \nJ & \pTjet[GeV] & \meff\ [GeV] & \met\ [GeV]\\
\cline{2-7}
& \multicolumn{6}{c}{other requirements}\\
\hline\hline
\multirow{2}{*}{VRWZ4j}
& $=3^*$ & $=0$ & $\geq 4$ & $> 25$ & $[600, 1500]$ & $[30,250]$ \\
\cline{2-7}
& \multicolumn{6}{c}{$\MetOverMeff < 0.2$,\quad$81<\mSFOS <\SI{101}{GeV}$} \\
\hline
\multirow{2}{*}{VRWZ6j}
& $=3^*$ & $=0$ & $\geq 6$ & $> 25$ & $[400, 1500]$ & $[30,250]$  \\
\cline{2-7}
& \multicolumn{6}{c}{$\MetOverMeff < 0.15$,\quad$81<\mSFOS <\SI{101}{GeV}$} \\
\hline
\multirow{2}{*}[-1.3em]{VRTTV}
& $\geq 2$ & $\geq 1$ & $\geq 3$ & $> 40$ & $[600, 1500]$ & $[30,250]$ \\
\cline{2-7}
& \multicolumn{6}{c}{
\begin{tabular}{@{}c@{}}
$\pt>\SI{30}{GeV}$ for the two leading-\pt same-sign leptons, \\
$\Delta R>1.1$ between the leading-\pt lepton and any jet, \\
$\HtBJetsOverHTJets > 0.4$,\quad$\MetOverMeff > 0.1$
\end{tabular}} \\
\hline
\multirow{2}{*}[-0.8em]{VRTTV1b6j}
& $\geq 2$ & $\geq 1$ & $\geq 6$ & $>40$ & $<1500$ & $[30,250]$ \\
\cline{2-7}
& \multicolumn{6}{c}{
\begin{tabular}{@{}c@{}}
$\pt>\SI{30}{GeV}$ for the two leading-\pt same-sign leptons, \\
$\MetOverMeff < 0.15$
\end{tabular}} \\
\hline
\multirow{2}{*}{VRTTW}
& $=2^*$ ($\mu^\pm\mu^\pm$) & $\geq 2$ & $\geq 2$ & $>25$ & $<1500$ & $[30,250]$ \\
\cline{2-7}
& \multicolumn{6}{c}{both leptons with $\pt>\SI{25}{GeV}$, one with $\pt>\SI{40}{GeV}$}\\
\hline
\multirow{2}{*}{VRTTW3j}
& $=2^*$ ($e^\pm\mu^\pm$) & $\geq 2$ & $\geq 3$ & $> 25$   & $<1500$ & $[30,250]$ \\
\cline{2-7}
& \multicolumn{6}{c}{both leptons with $\pt>\SI{25}{GeV}$}\\
\hline\hline
\multicolumn{7}{l}{$^*$: additional baseline leptons are not allowed}\\
\hline\hline
\end{tabular}
\end{table}
 
\begin{figure}[t!]
\begin{center}
\includegraphics[width=1\textwidth]{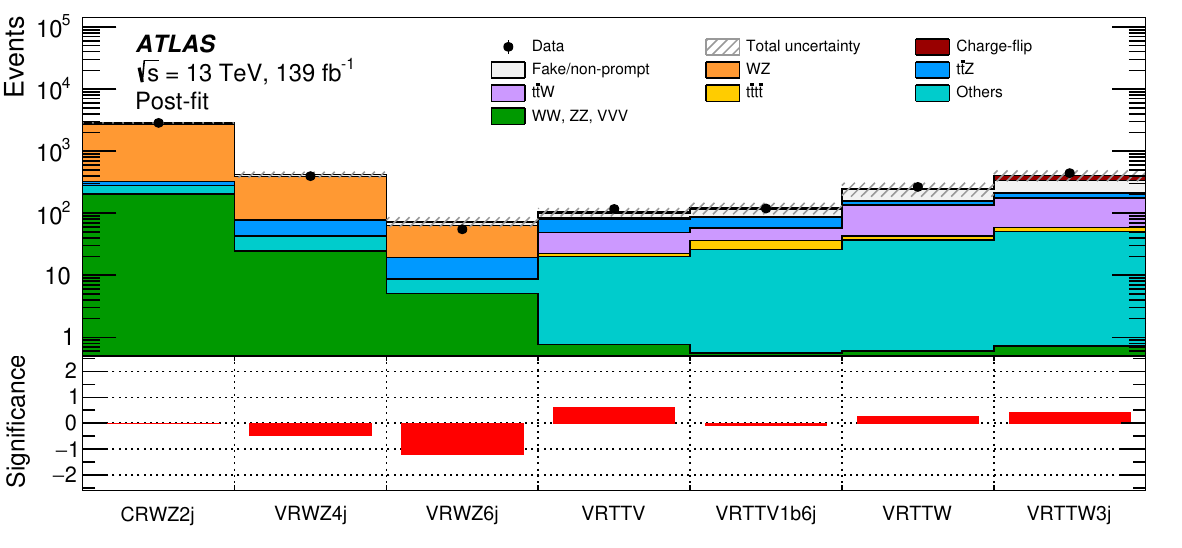}
\end{center}
\caption{Data and post-fit background comparison in CRWZ2j and the VRs.
The total uncertainties in the expected event yields are shown as the hatched bands.
The bottom panel shows the significance~\cite{ATL-PHYS-PUB-2020-025} quantifying the deviation of the observed yields
from the background expectation.
}
\label{fig:summary_VRyields}
\end{figure}

The $WZ$\,+\,jets, $\ttbar W$ and $\ttbar\ell^+\ell^-$ background predictions
are checked in the validation regions (VRs) defined in Table~\ref{tab:VRs_defs}.
The purity in the targeted SM process is expected to range from 70\% to 80\% for the two VRWZ regions,
and from 30\% to 55\% for the four VRTTV/VRTTW regions,
while the expected contributions from SUSY signal processes remain small.
The VRs are designed to be orthogonal to the SRs,
with the exception of VRWZ4j and VRWZ6j, which
partially overlap with four of the SRGGWZ and SRSSWZ regions.
However, for each of those, the expected number shared SM background events does not exceed
3\% of the total number of events in the VR.
The background predictions and the observed numbers of events in CRWZ2J and in the VRs
are illustrated in Figure~\ref{fig:summary_VRyields}.
The data and the background expectation in all VRs agree within uncertainties.
The largest tension occurs in VRWZ6j, but it was checked that in this region the level of agreement doesn't become poorer as a function of \meff or \met.
 
\section{Systematic uncertainties}
\label{sec:systematics}

The predicted background yields in the SRs are affected by several sources of systematic and statistical uncertainty. The systematic uncertainties are grouped into experimental and theoretical ones, as well as those arising from the data-driven methods described above, normalisation, and MC statistics.
 
The experimental uncertainties come from the possible differences between the data and simulations in elements of this analysis and the uncertainties in the data taken from the operating detectors.
They are related to luminosity, \pileup, triggers, and the reconstructed objects.
For luminosity, a 1.7\% relative uncertainty~\cite{LUCID2} is applied.
For \pileup, the uncertainty is computed by increasing and decreasing the weights associated with the mean number of simultaneous interactions by 4\%.
For leptons, uncertainties are computed for reconstruction efficiencies~\cite{EGAM-2018-01}, identification efficiencies~\cite{MUON-2018-03, PERF-2017-01}, isolation efficiencies~\cite{ATLAS-CONF-2016-024}, energy scales and resolutions~\cite{EGAM-2018-01, MUON-2022-01}, and trigger efficiencies~\cite{TRIG-2018-05, TRIG-2018-01, TRIG-2019-01} using various methods.
For jets, uncertainties are considered for the jet vertex tagger (JVT)~\cite{ATL-PHYS-PUB-2019-026}, jet energy scale (JES)~\cite{PERF-2016-04} and jet energy resolution (JER)~\cite{JETM-2018-05}, and flavour tagging~\cite{FTAG-2018-01,FTAG-2020-08,ATLAS-CONF-2018-006}.
For the \met, uncertainties are estimated by propagating the uncertainties in the energy and momentum scale of each of the objects entering the calculation, and the uncertainties in the soft term's resolution and scale~\cite{ATLAS-CONF-2018-023}.
 
The theoretical uncertainties come from the MC modelling of the relevant SM and SUSY processes, including cross sections, choice of scales, the PDF and \alphas. The theoretical uncertainties from the dominant background processes in the signal regions, such as $WZ$, $W^{\pm}W^{\pm}$, and $t\bar{t}V$ (as well as $\ttbar\ttbar$ for RPV signal regions), are computed in detail.
Uncertainties from the choice of renormalisation scales, factorisation scales, resummation scales, merging scales and the recoil schemes, are evaluated by using the varied scales instead of the chosen scales and comparing the results.
The impact of the choice of PDF is evaluated by symmetrising the event count variations seen when using the \MMHT~\cite{Harland-Lang:2014zoa}, \CT[14]~\cite{Dulat:2015mca}, and \NNPDF~\cite{Ball:2012cx} PDF sets.
The total cross-section uncertainty is not applied to the $WZ$ process because this process is normalised to data in the CRWZ2j control region.
For the remaining rarer processes listed in Table~\ref{tab:generators}, an overall 50\% uncertainty is assigned.
While some of the SRs are divided into several bins of a particular observable,
the theory uncertainties are assumed to affect all bins uniformly.
The impact of this simplification upon exclusion limits presented in the following sections was found to be acceptably small.
 
\begin{figure}[!htb]
\centering
\includegraphics[width=0.9\textwidth]{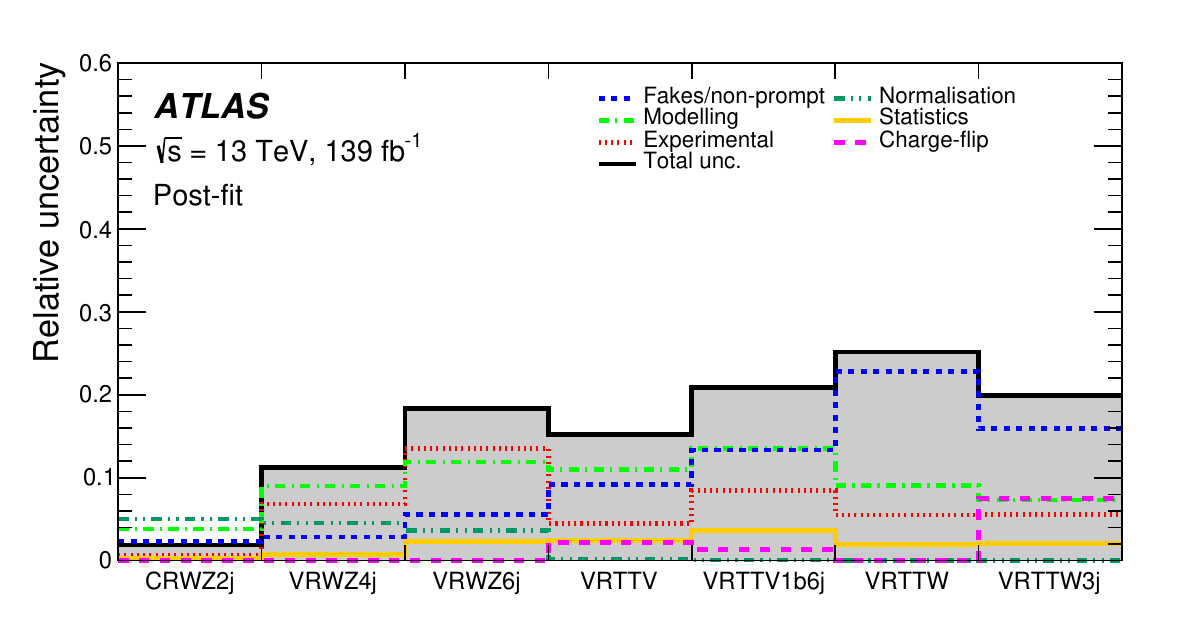}
 
\caption{ Relative contributions from different categories of uncertainties in CRWZ2j and the VRs.
Correlations between different sources of systematic uncertainty are considered, so the total uncertainty does not necessarily match the quadrature summation of the components.
}
\label{fig:vr_syst_fig1_postfit}
\end{figure}

\begin{figure}
\centering
\includegraphics[width=0.9\textwidth]{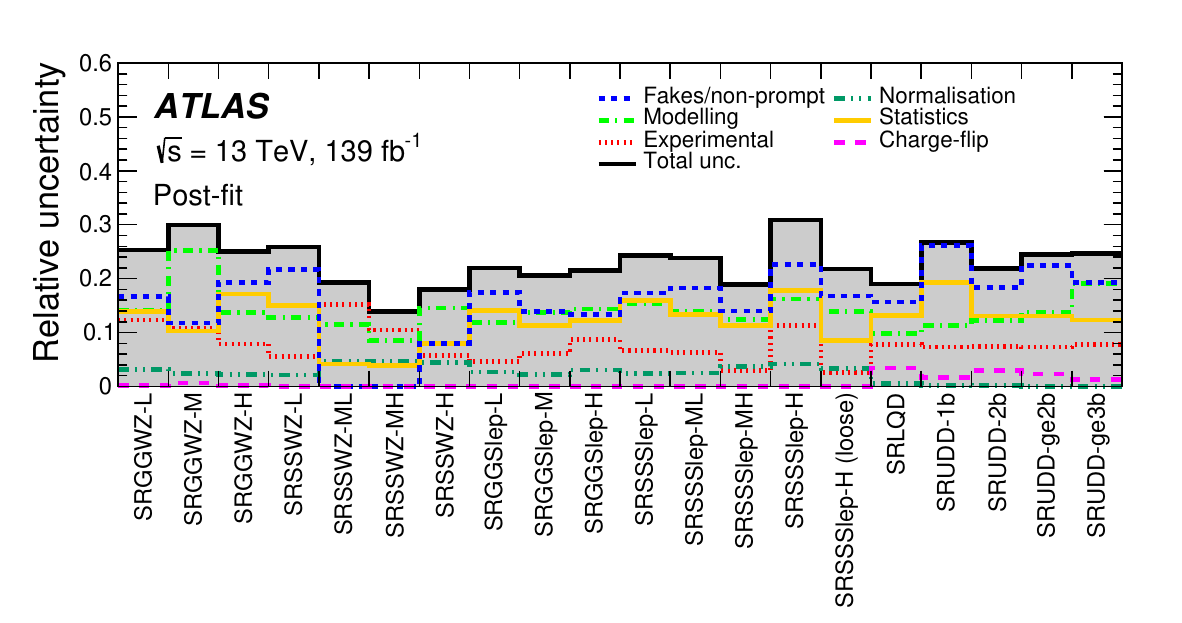}
\caption{Relative contributions from different categories of uncertainties in the signal regions.
Correlations between different sources of systematic uncertainty are considered, so the total uncertainty does not necessarily match the quadrature summation of the components.}
\label{fig:SR_sys}
\end{figure}

For the purpose of defining correlations,
the different processes are grouped into nine categories:
$WZ$, $t\bar{t}W$,  $t\bar{t}\ell\ell$, $\ttbar\ttbar$, other multiboson processes, other processes with top quarks, F/NP leptons, charge-flip electrons, and SUSY signal.
Within each group, every source of uncertainty is treated as fully correlated,
including across the different regions involved in the simultaneous fit.
Experimental uncertainties are also treated as correlated between the different groups,
while all other uncertainties are assumed to be uncorrelated.
The total uncertainty and the separate contributions from each source are shown in Figure~\ref{fig:vr_syst_fig1_postfit} and Figure~\ref{fig:SR_sys} for CR/VRs and SRs separately.
The total uncertainties vary from 2\% (CRWZ2j) to 31\% (SRSSSlep-H). The dominant contributions come from the F/NP-lepton background estimate, and the MC statistics in the SRs.
The total uncertainty in CRWZ2j is smaller than the sum in quadrature of the individual components because of the anti-correlation between the normalisation and modelling uncertainties.
 
\section{Statistical analysis}
\label{sec:statistical}

 
The expected SM backgrounds are determined with a profile likelihood fit~\cite{Baak:2014wma}, referred to as a background-only fit.
The fit strategy differs between the RPC and RPV searches.
For the RPC searches, the background-only fit uses the observed event yield in CRWZ2j as a constraint to adjust the normalisation of the $WZ$\,+\,jets background assuming that no signal is present.
The inputs to the background-only fit include the number of events observed in CRWZ2j, and the number of events predicted in CRWZ2j, and in the SR(s) of interest, for all background processes.
Both the observed and predicted numbers of events are described by Poisson statistics.
The systematic uncertainties are included in the fit as nuisance parameters.
They are constrained by Gaussian distributions with widths corresponding to the sizes of the uncertainties and are treated as correlated, when appropriate, between the various regions.
The product of the various probability density functions forms the likelihood, which the fit maximises by
adjusting the normalisation of the $WZ$\,+\,jets background and the nuisance parameters.
For the RPV searches, CRWZ2j is not included in the fit, and the likelihood fit is used just to constrain the nuisance parameters associated with the systematic uncertainties.
In both cases, the results of the background-only fit are used to test the compatibility of the observed data and the background estimates in the  SRs.
 
In the absence of a significant excess over the SM expectation, two levels of interpretation are provided for BSM physics scenarios:
model-independent exclusion limits and model-dependent exclusion limits set on the SUSY benchmark models illustrated in Figure~\ref{fig:signals}.
The  CL$_{\textrm{s}}$  method~\cite{Junk:1999kv,Read:2002hq} is used to derive the confidence level (CL) of the exclusion for a particular signal model.
A  signal model with a  CL$_{\textrm{s}}$  value below 0.05 is excluded at 95\% CL.

Model-independent exclusion fits are used to set 95\% CL upper limits on the possible BSM contributions to the SRs.
This fit proceeds in the same way as the background-only fit, with CRWZ2j (for RPC searches) and the SRs both participating in a simultaneous likelihood fit,
and the likelihood function including an additional parameter-of-interest to describe the potential signal contribution.
Signal contamination in CRWZ2j is assumed to be zero.
The hypothesis tests are performed for each of the SRs independently.  The limits were evaluated using pseudo-experiments.
 
Model-dependent exclusion fits are used to set 95\% CL exclusion limits on the masses of  gluinos and squarks for the SUSY benchmark models considered in this paper.
The fit proceeds similarly to the model-independent fit, except that both the signal yield in the SRs and the
signal contamination in CRWZ2j expected from the model are taken into account, and the SRs are usually binned.
For the models shown in Figures \ref{fig:models_SSslep}  and \ref{fig:RPV_UDD}, only the SRs with  the best expected sensitivity are considered.
Table~\ref{tab:fitConfig} shows the SRs used for each benchmark model and the fitted observable in each SR.
The observable providing the best sensitivity for each SR is chosen as the fitted variable.
The binning of each observable was optimised to provide the best sensitivity for the benchmark model of interest, while keeping enough events in each bin of the fitted SR.
This multi-bin approach was found to enhance the sensitivity for all the SUSY scenarios considered in this search.
In this model-dependent fit, the CL$_{\textrm{s}}$ value is computed using the asymptotic approximation~\cite{Cowan:2010js}.  For some selected points a comparison with \enquote{toy} experiments was performed, and the exclusion mass limits are overestimated by a maximum of 2\%.

\begin{table}[htp]
\caption{Fit configuration used to obtain the exclusion limits for each benchmark model.  The targeted signal model is shown in the first column. The second and third columns show the signal regions and the  fitted variable in each signal region, respectively.  A statistical combination of signal regions is represented by  the symbol  \enquote{\&},  while \enquote{||} means that for each point of the $\{m_{\tilde g (\tilde q)},m_{\tilde\chi_1^0}\}$ parameter space, the signal region with the best expected sensitivity is chosen.}
\label{tab:fitConfig}
\resizebox{0.99\textwidth}{!}{
\centering
\begin{tabular}{c|c|c}
\hline
\hline
Model  &  Signal region(s)  &  Variable  \\
\hline
\hline
 
$\tilde g \rightarrow qq^\prime WZ \tilde\chi_1^0 $  & \multirow{2}{*}{SRGGWZ-L ||   SRGGWZ-M ||  SRGGWZ-H} & \multirow{2}{*}{single-bin, \meff, single-bin}   \\
Figure~\ref{fig:models_GGwz}  &    &   \\
\hline

$\tilde q \rightarrow q^\prime WZ \tilde\chi_1^0 $  &   \multirow{2}{*}{SRSSWZ-L || SRSSWZ-ML || SRSSWZ-MH || SRSSWZ-H}  & \multirow{2}{*}{\met, \met, \meff, \meff }  \\
 
Figure~\ref{fig:models_SSwz}  &    &   \\
\hline
 
$\tilde g \rightarrow q\bar q(\ell\ell/\nu\nu) \tilde\chi_1^0 $ &  \multirow{2}{*}{SRGGSlep-L || SRGGSlep-M || SRGGSlep-H}  &  \multirow{2}{*}{ \MetOverHTLep,  \met,  \met}  \\
 
Figure~\ref{fig:models_GGslep} &    &   \\
\hline

$\tilde q \rightarrow   q(\ell\nu/\ell\ell/\nu\nu)\tilde\chi_1^0 $  &  \multirow{2}{*}{SRSSSlep-L || SRSSSlep-ML || SRSSSlep-MH || SRSSSlep-H (loose)} & \multirow{2}{*}{ \meff\ }   \\
Figure~\ref{fig:models_SSslep}  &    &   \\
 
\hline
$\tilde g \rightarrow q\bar q \tilde\chi_1^0 $, $\tilde\chi_1^0 \rightarrow \ell qq$  &   \multirow{2}{*}{SRLQD}   &   \multirow{2}{*}{  \meff\ }  \\
 
Figure~\ref{fig:models_GGrpv} &    &   \\
\hline
 
$\tilde g \rightarrow \tilde t \bar{t}$, $\tilde t \rightarrow \bar{b}\bar{d}$ &  \multirow{2}{*}{SRUDD-1b   \&  SRUDD-ge2b}  &    \multirow{2}{*}{ \HTJets\ }      \\

Figure~\ref{fig:RPV_UDD} &    &   \\
\hline
\hline
 
\end{tabular}
}
\end{table}


\section{Results}
\label{sec:result}

\begin{figure}[htb!]

\begin{center}
 
\includegraphics[width=1\textwidth]{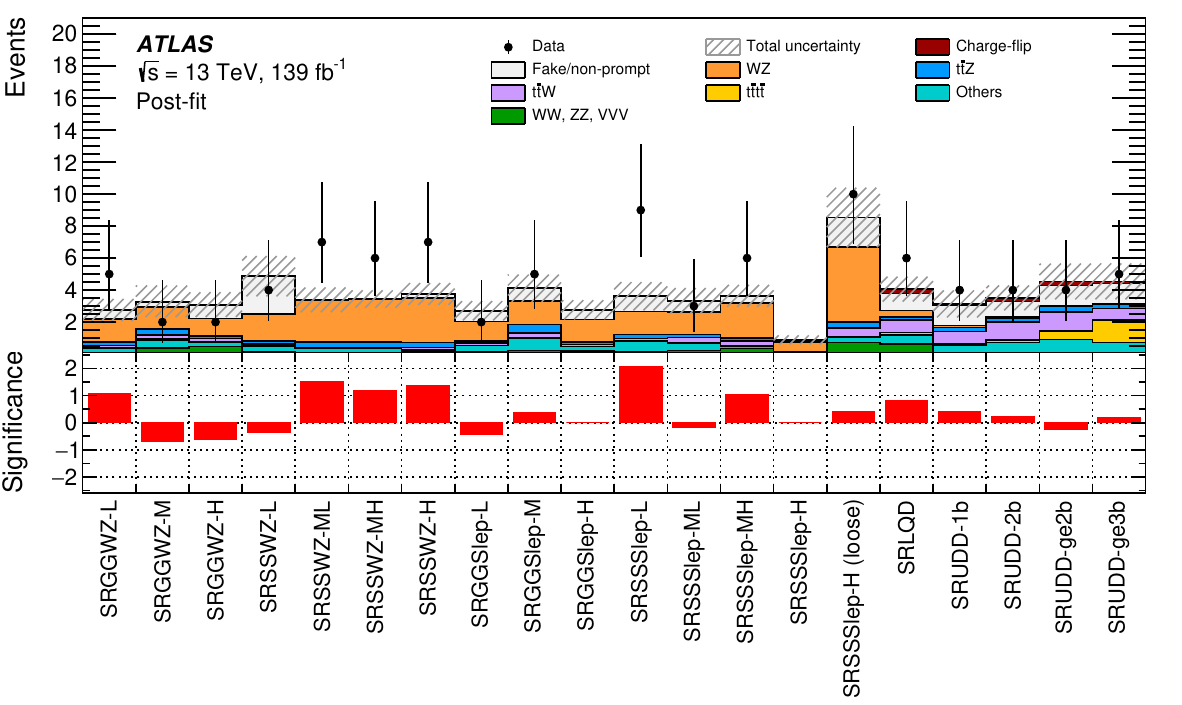}
 
\end{center}
 
\caption{Data and post-fit background comparison in all SRs. The total uncertainties in the expected event yields are shown as the hatched bands.  The bottom panel shows the significance~\cite{ATL-PHYS-PUB-2020-025} quantifying the deviation of the observed yields from the background expectation.}
\label{fig:summary_yields}
\end{figure}

\begin{figure}[htb!]
 
\begin{center}

\begin{subfigure}{0.37\textwidth}
\includegraphics[width=1\textwidth]{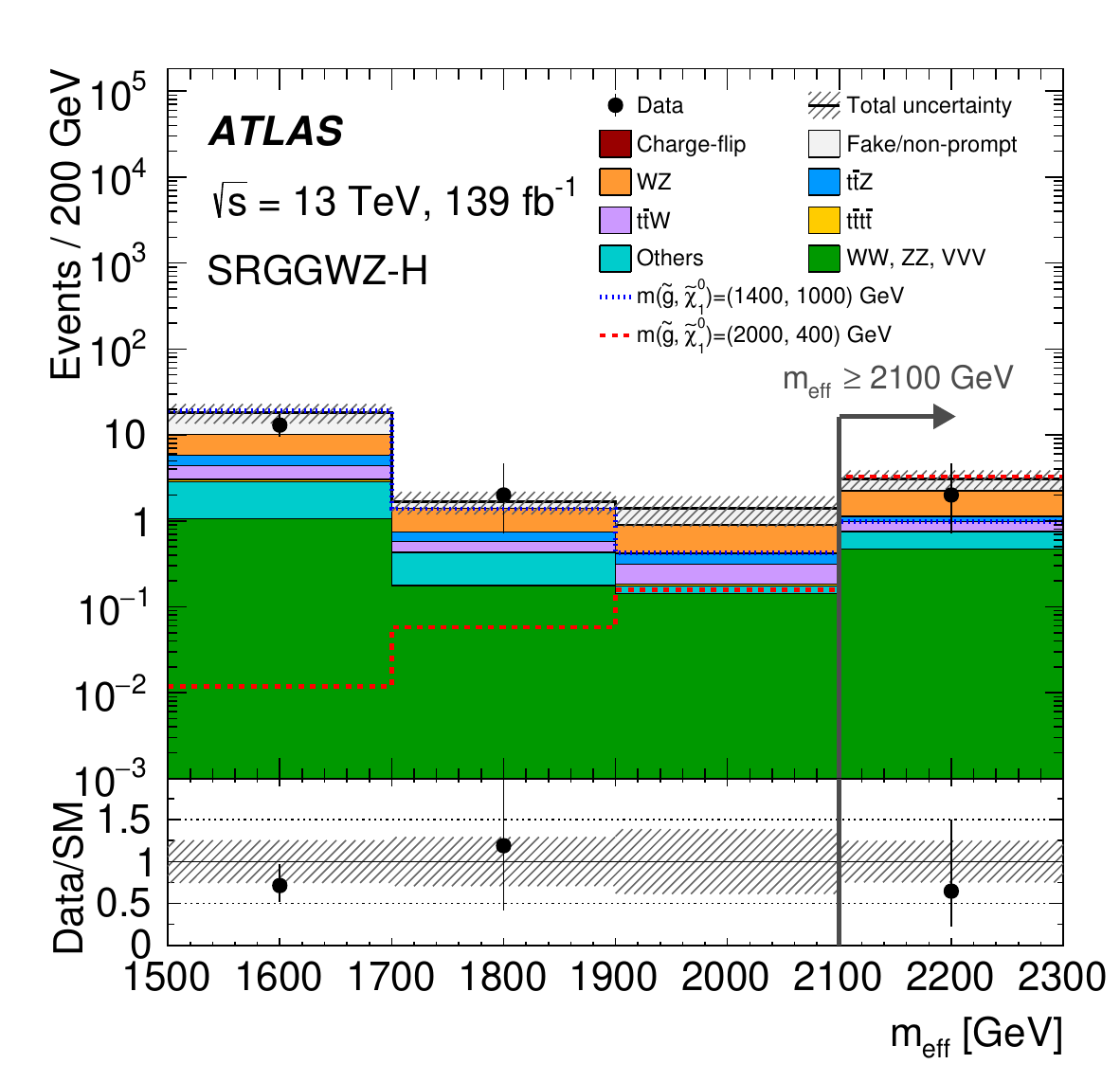}
\caption{SRGGWZ-H}
\end{subfigure}
\begin{subfigure}{0.37\textwidth}
\includegraphics[width=1\textwidth]{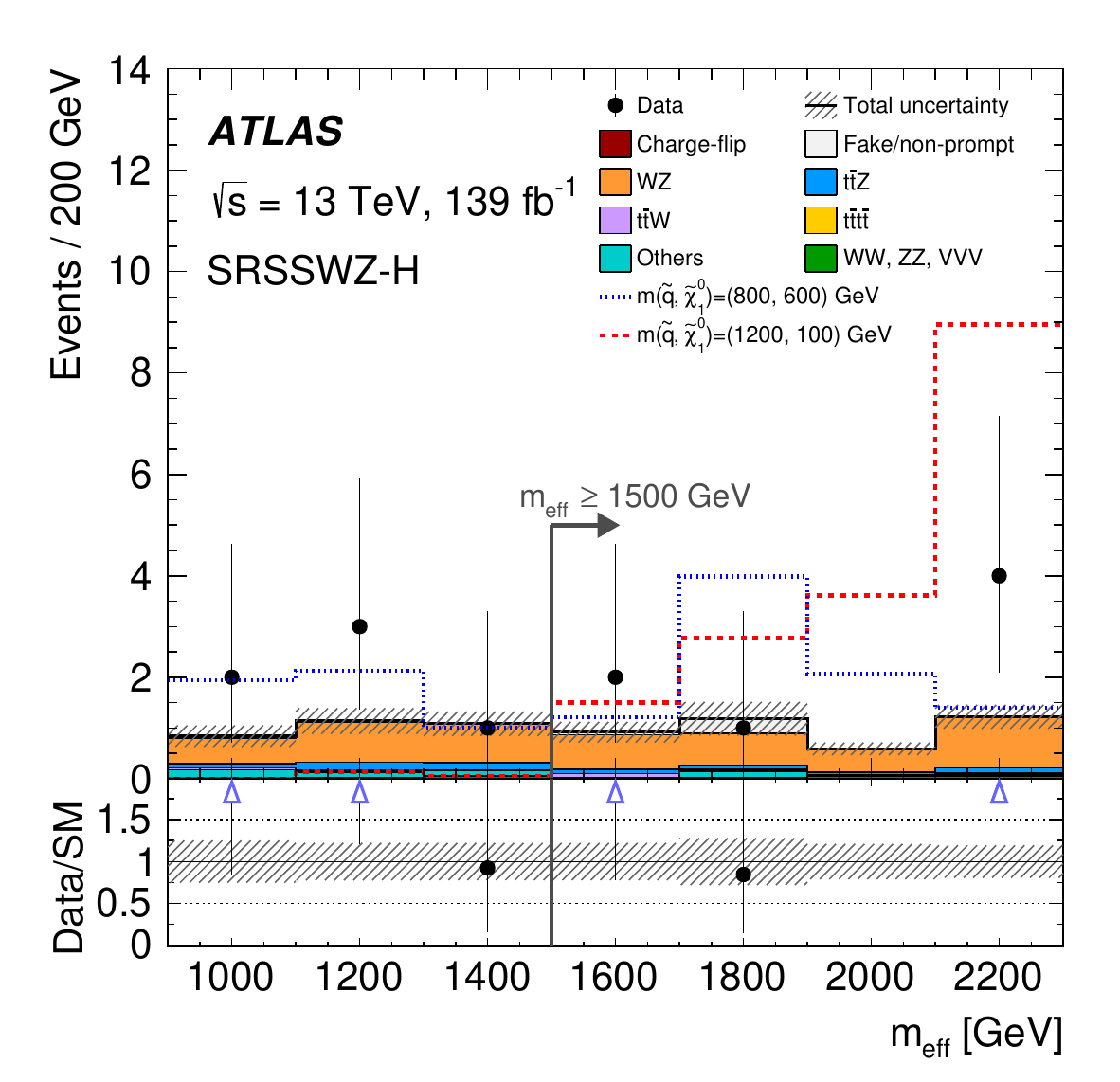}
\caption{ SRSSWZ-H}
\end{subfigure}

\vspace{0.15in}
 
\begin{subfigure}{0.37\textwidth}
\includegraphics[width=1\textwidth]{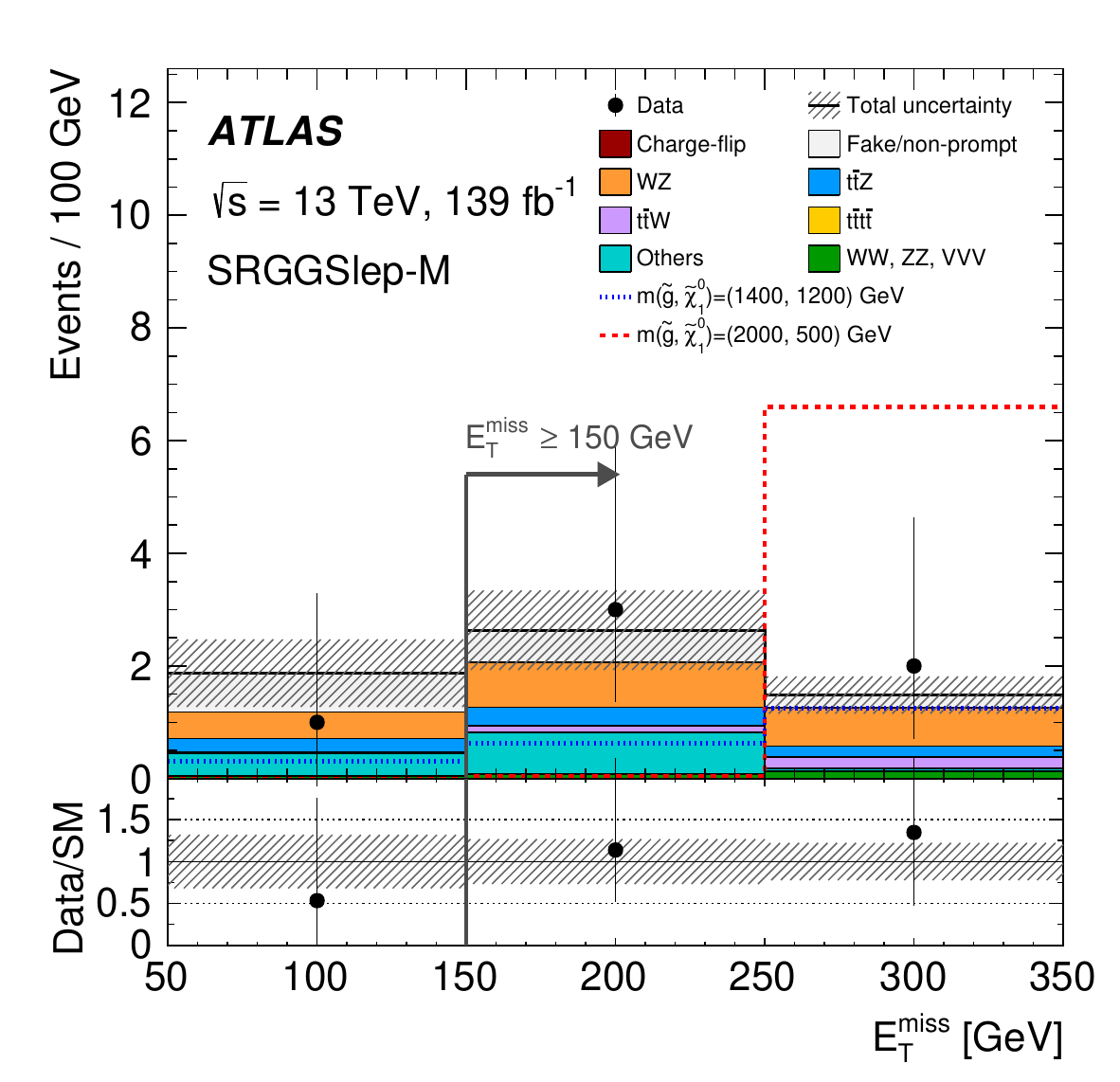}
\caption{SRGGSlep-M}
\end{subfigure}
\begin{subfigure}{0.37\textwidth}
\includegraphics[width=1\textwidth]{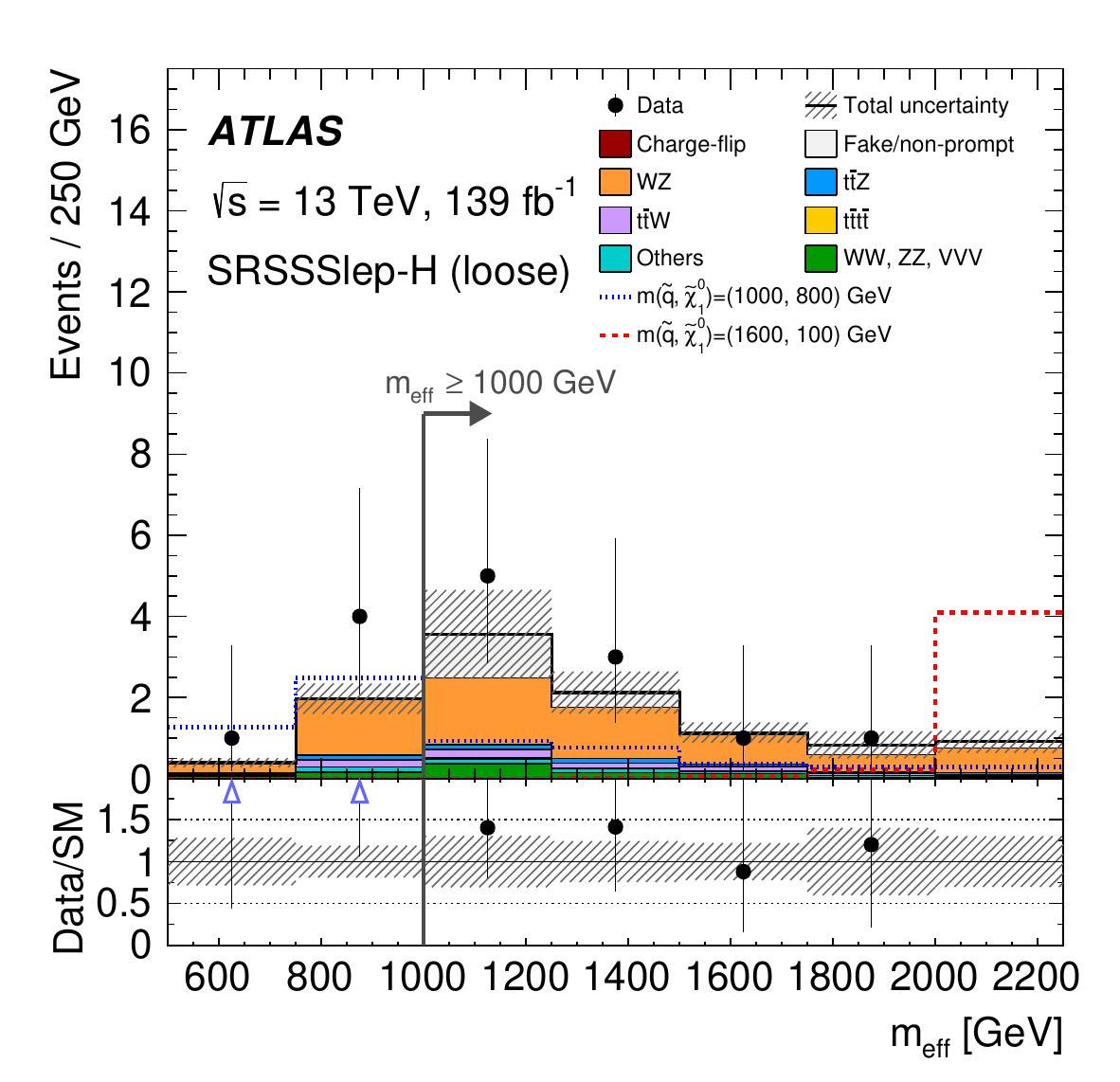}
\caption{SRSSSlep-H (loose)}
\end{subfigure}
 
\vspace{0.15in}
\begin{subfigure}{0.37\textwidth}
\includegraphics[width=1\textwidth]{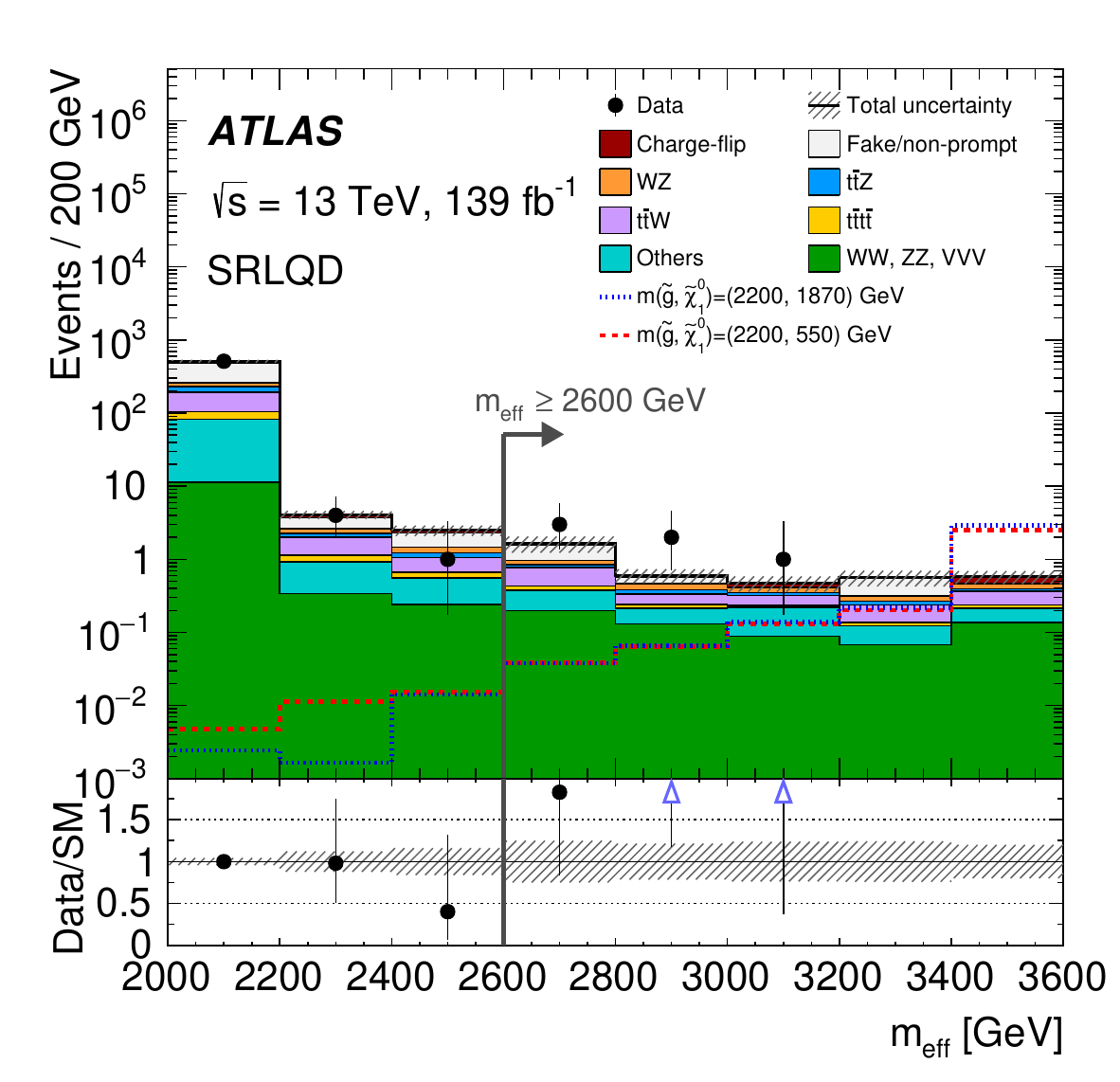}
\caption{SRLQD}
\end{subfigure}
\begin{subfigure}{0.37\textwidth}
\includegraphics[width=1\textwidth]{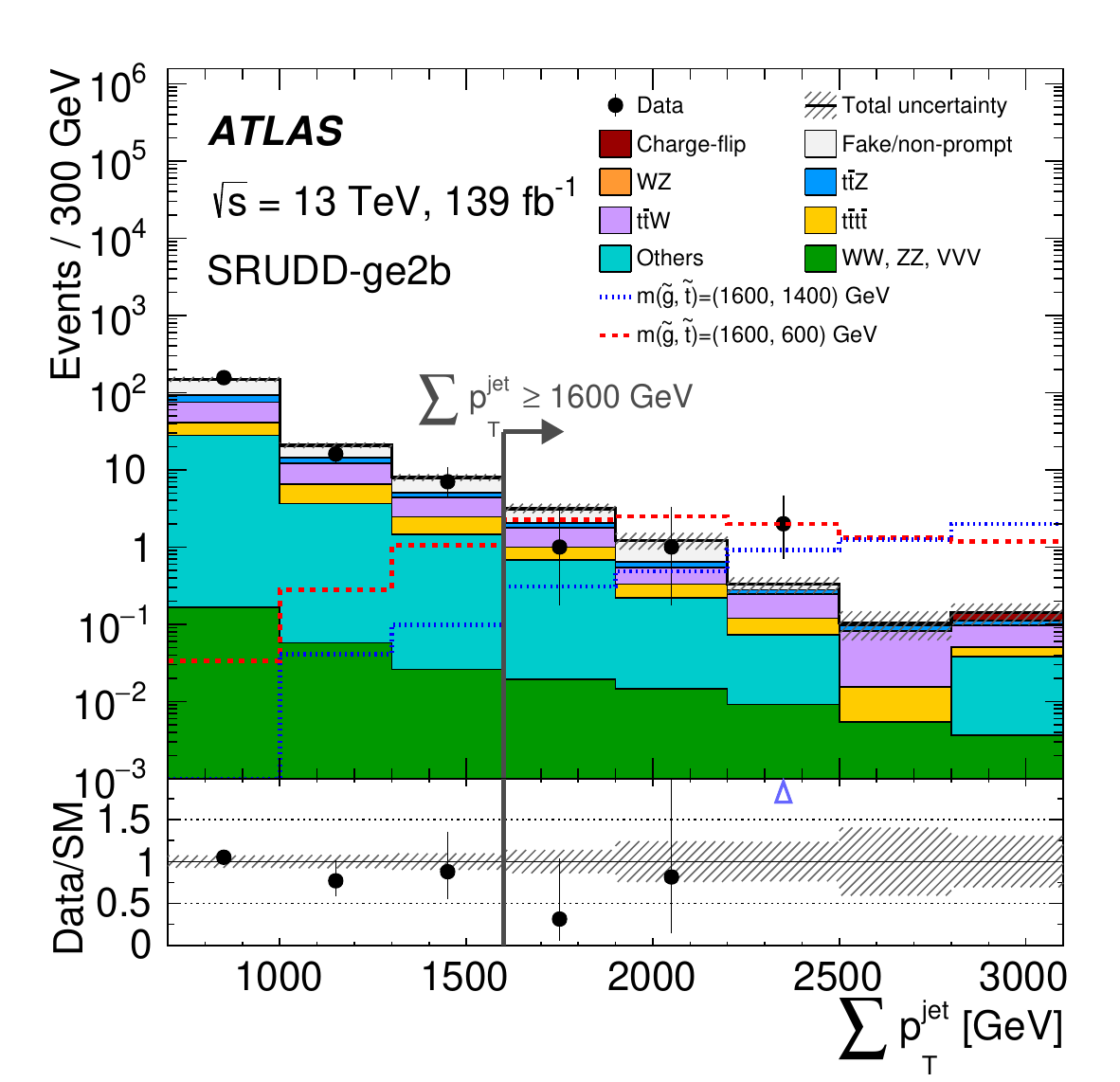}
\caption{SRUDD-ge2b}
\end{subfigure}
 
\end{center}
 
\caption{ Distributions of the data and estimated background after the background-only fit for the signal regions (a) SRGGWZ-H, (b) SRSSWZ-H, (c) SRGGSlep-M, (d) SRSSSlep-H (loose), (e) SRLQD, and (f) SRUDD-ge2b.
All SR selections but the one on the quantity shown are applied.
The line with an arrow indicates the  requirement used in that signal region.
Distributions for two signal hypotheses for the model of interest are also shown.
The bins displayed above the signal region requirement correspond to the binning used in the exclusion fit.
All uncertainties are included in the uncertainty band.
Overflow (underflow) events, where present, are included in the last (first) bin.
\label{fig:N1sec10}}
\end{figure}

 
The observed number of events in each SR along with the background expectations and uncertainties are summarised in Figure~\ref{fig:summary_yields}, and shown in detail in Tables~\ref{tab:observed_sr_yields_table1_postfit_RpcWZ}--\ref{tab:observed_sr_yields_table1_postfit_Rpv}.
The  background prediction corresponds to the estimate after the background-only fit described in Section~\ref{sec:statistical}.
The overall excess (less than 2 standard deviations) observed in  SRSSWZ-ML, SRSSWZ-MH, and SRSSWZ-H, is due to the overlap among  these regions, which have three data events in common.
Overlap among the remaining SRs is also observed, but in a smaller proportion.
The contribution from $WZ$\,+\,jets dominates in the SRs with no $b$-jets, while the  production of fake/non-prompt leptons and processes involving the top quark dominate in the SRs where the veto on $b$-jets is not applied.
No significant excess of events above the SM prediction is observed in any of the SRs.
The highest significance corresponds to SRSSSlep-L with 2.2 standard deviations
(with the calculation of Table~\ref{tab:limits_mi}).
The distributions of the most discriminating variable for some of the SRs, with the signal region requirement on the displayed variable removed, are shown in Figure~\ref{fig:N1sec10}.

\begin{table}[htb!]
\begin{center}
\caption{The number of observed data events and expected background contributions in signal regions defined for the  RPC models shown in Figures~\ref{fig:models_GGwz} and \ref{fig:models_SSwz}. Backgrounds shown with a \enquote{$-$} do not contribute to that region. The displayed yields include all statistical and systematic uncertainties. The individual uncertainties can be correlated or anticorrelated and therefore do not necessarily add in quadrature to equal the total uncertainty.
}
\label{tab:observed_sr_yields_table1_postfit_RpcWZ}
\setlength{\tabcolsep}{0.0pc}{
\scriptsize
\begin{tabular*}{\textwidth}{@{\extracolsep{\fill}}lccccccc}
\noalign{\smallskip}\hline\hline\noalign{\smallskip}
 
&    SRGGWZ-L    &    SRGGWZ-M    &    SRGGWZ-H    &    SRSSWZ-L    &    SRSSWZ-ML    &    SRSSWZ-MH    &    SRSSWZ-H    \\[-0.05cm]
 
\noalign{\smallskip}\hline\noalign{\smallskip}
 
Observed    &    5    &    2    &    2    &    4    &    7    &    6    &    7    \\
 
\noalign{\smallskip}\hline\hline\noalign{\smallskip}
 
Total background    &    2.8 $\pm$ 0.7    &    3.3 $\pm$ 1.0    &    3.1 $\pm$ 0.8    &    4.9 $\pm$ 1.3    &    3.4 $\pm$ 0.8    &    3.4 $\pm$ 0.6    &    3.8 $\pm$ 0.7  \\
 
\noalign{\smallskip}\hline\noalign{\smallskip}
 
$WZ$     &    1.4 $\pm$ 0.4    &    1.3 $\pm$ 0.8    &    1.1 $\pm$ 0.4    &    1.7 $\pm$ 0.6    &    2.6 $\pm$ 0.7    &    2.7 $\pm$ 0.5    &    2.8 $\pm$ 0.6    \\
 
$ZZ, \WWSS, VVV$     &    0.09 $\pm$ 0.05    &    0.37 $\pm$ 0.11    &    0.47 $\pm$ 0.11    &    0.14 $\pm$ 0.07    &    0.09 $\pm$ 0.06    &    0.11 $\pm$ 0.07    &    0.15 $\pm$ 0.07    \\
 
$t\bar{t}W$     &    0.15 $\pm$ 0.09    &    0.26 $\pm$ 0.12    &    0.22 $\pm$ 0.07    &    0.12 $\pm$ 0.07    &    $<0.05$    &   $<0.05$    &    0.14 $\pm$ 0.04    \\
 
$t\bar{t}Z$     &    0.24 $\pm$ 0.08    &    0.38 $\pm$ 0.22    &    0.15 $\pm$ 0.10    &    0.18 $\pm$ 0.10    &    0.38 $\pm$ 0.14    &    0.38 $\pm$ 0.12    &    0.31 $\pm$ 0.10    \\
 
$\ttbar\ttbar$     &    0.02 $\pm$ 0.01    &    0.04 $\pm$ 0.02    &    $<0.02$    &    $<0.02$ &    $<0.02$     &    $<0.02$     &    $<0.02$    \\
 
Other SM processes   &    0.26 $\pm$ 0.18    &    0.53 $\pm$ 0.28    &    0.28 $\pm$ 0.15    &    0.36 $\pm$ 0.20    &    0.28 $\pm$ 0.31    &    0.25 $\pm$ 0.21    &    0.12 $\pm$ 0.10    \\
 
Fake/non-prompt     &    0.56 $\pm$ 0.46    &    0.3 $\pm$ 0.4    &    0.9 $\pm$ 0.6    &    2.4 $\pm$ 1.1    &    $<0.3$   &    $<0.3$    &    0.24 $\pm$ 0.27    \\
 
Charge-flip     &    $<0.02$    &    0.03 $\pm$ 0.02    &    $<0.02$     &    --    &    --    &   --    &    --    \\
 
\noalign{\smallskip}\hline\hline\noalign{\smallskip}
\end{tabular*}
}
\end{center}
\end{table}
 
\begin{table}[htb!]
\begin{center}
\caption{The number of observed data events and expected background contributions  in signal regions defined for the  RPC models  shown in Figures~\ref{fig:models_GGslep} and \ref{fig:models_SSslep}. Backgrounds shown with a \enquote{$-$} do not contribute to that region. The displayed yields include all statistical and systematic uncertainties. The individual uncertainties can be correlated or anticorrelated and therefore do not necessarily add in quadrature to equal the total uncertainty.}
\setlength{\tabcolsep}{0.0pc}{
\resizebox*{\textwidth}{!}{
\small
\begin{tabular*}{1.3\textwidth}{@{\extracolsep{\fill}}lcccccccc}
\noalign{\smallskip}\hline\hline\noalign{\smallskip}
 
&    SRGGSlep-L    &    SRGGSlep-M    &    SRGGSlep-H    &    SRSSSlep-L    &    SRSSSlep-ML    &    SRSSSlep-MH    &    SRSSSlep-H    &    SRSSSlep-H (loose) \\[-0.05cm]
 
\noalign{\smallskip}\hline\noalign{\smallskip}
 
Observed    &    2    &    5    &    0    &    9    &    3    &    6    &    0    &    10    \\
 
\noalign{\smallskip}\hline\hline\noalign{\smallskip}
 
Total background    &    2.7 $\pm$ 0.6    &    4.1 $\pm$ 0.9    &    2.8 $\pm$ 0.6    &    3.6 $\pm$ 0.9    &    3.3 $\pm$ 0.8    &    3.6 $\pm$ 0.7    &    0.89 $\pm$ 0.28    &    8.5 $\pm$ 1.9 \\
 
\noalign{\smallskip}\hline\noalign{\smallskip}
 
$WZ$     &    1.19 $\pm$ 0.27    &    1.5 $\pm$ 0.4    &    1.4 $\pm$ 0.4    &    1.4 $\pm$ 0.4    &    1.4 $\pm$ 0.4    &    2.2 $\pm$ 0.4    &    0.61 $\pm$ 0.17    &    4.7 $\pm$ 1.1    \\
 
$ZZ, \WWSS, VVV$     &    0.14 $\pm$ 0.08    &    0.21 $\pm$ 0.11    &    0.18 $\pm$ 0.09    &    0.13 $\pm$ 0.07    &    0.23 $\pm$ 0.12    &    0.34 $\pm$ 0.17    &    0.04 $\pm$ 0.03    &    0.7 $\pm$ 0.4    \\
 
$t\bar{t}W$     &    0.15 $\pm$ 0.08    &    0.31 $\pm$ 0.09    &    0.11 $\pm$ 0.05    &    0.14 $\pm$ 0.08    &    0.35 $\pm$ 0.08    &    0.30 $\pm$ 0.07    &    $<0.05$    &    0.55 $\pm$ 0.29    \\
 
$t\bar{t}Z$     &    0.15 $\pm$ 0.08    &    0.53 $\pm$ 0.17    &    0.15 $\pm$ 0.09    &    0.23 $\pm$ 0.13    &    0.19 $\pm$ 0.08    &    0.16 $\pm$ 0.09    &    0.05 $\pm$ 0.03    &    0.36 $\pm$ 0.22    \\
 
$\ttbar\ttbar$     &    $<0.02$    &    0.02 $\pm$ 0.01    &    0.02 $\pm$ 0.01    &    $<0.02$   &    $<0.02$    &    $<0.02$    &    $<0.02$    &    0.02 $\pm$ 0.01    \\
 
Other SM processes     &    0.39 $\pm$ 0.23    &    0.8 $\pm$ 0.4    &    0.29 $\pm$ 0.18    &    0.7 $\pm$ 0.4    &    0.44 $\pm$ 0.25    &    0.17 $\pm$ 0.09    &    0.02 $\pm$ 0.03   &    0.34 $\pm$ 0.18    \\
 
Fake/non-prompt     &    0.7 $\pm$ 0.5    &    0.8 $\pm$ 0.6    &    0.6 $\pm$ 0.4    &    1.0 $\pm$ 0.6    &    0.7 $\pm$ 0.6    &    0.5 $\pm$ 0.5    &    0.14 $\pm$ 0.20    &    1.9 $\pm$ 1.4    \\
 
Charge-flip     &    --    &    --   &   --    &    --    &    --   &    --    &    --   &    --    \\
 
\noalign{\smallskip}\hline\hline\noalign{\smallskip}
\end{tabular*} }}
\end{center}
\end{table}

\begin{table}[htb!]
\begin{center}
\caption{The number of observed data events and expected background contributions  in signal regions defined for the  RPV models  shown in Figures~\ref{fig:models_GGrpv} and \ref{fig:RPV_UDD}. The displayed yields include all statistical and systematic uncertainties. The individual uncertainties can be correlated or anticorrelated and therefore do not necessarily add in quadrature to equal the total uncertainty. }
\label{tab:observed_sr_yields_table1_postfit_Rpv}
\setlength{\tabcolsep}{0.0pc}{
\scriptsize
\begin{tabular*}{1\textwidth}{@{\extracolsep{\fill}}lccccc}
\noalign{\smallskip}\hline\hline\noalign{\smallskip}
 
&    SRLQD    &    SRUDD-1b    &    SRUDD-2b    &    SRUDD-ge2b    &    SRUDD-ge3b    \\[-0.05cm]
 
\noalign{\smallskip}\hline\noalign{\smallskip}
 
Observed    &    6    &    4    &    4    &    4    &    5    \\
 
\noalign{\smallskip}\hline\hline\noalign{\smallskip}
 
Total background    &    4.1 $\pm$ 0.8    &    3.1 $\pm$ 0.8    &    3.5 $\pm$ 0.8    &    4.5 $\pm$ 1.1    &    4.6 $\pm$ 1.1  \\
 
\noalign{\smallskip}\hline\noalign{\smallskip}
 
$WZ$     &    0.40 $\pm$ 0.14    &    0.11 $\pm$ 0.06    &    0.10 $\pm$ 0.10    &    0.03 $\pm$ 0.02    &    $<0.02$    \\
 
$ZZ, \WWSS, VVV$     &    0.65 $\pm$ 0.22    &    0.06 $\pm$ 0.03    &    0.08 $\pm$ 0.05    &    0.05 $\pm$ 0.03    &    $<0.02$   \\
 
$t\bar{t}W$     &    0.78 $\pm$ 0.23    &    0.78 $\pm$ 0.24    &    1.11 $\pm$ 0.27    &    1.20 $\pm$ 0.30    &    0.7 $\pm$ 0.4    \\
 
$t\bar{t}Z$     &    0.20 $\pm$ 0.08    &    0.27 $\pm$ 0.17    &    0.26 $\pm$ 0.13    &    0.34 $\pm$ 0.19    &    0.24 $\pm$ 0.11    \\
 
$\ttbar\ttbar$     &    0.14 $\pm$ 0.09    &    0.10 $\pm$ 0.06    &    0.17 $\pm$ 0.11    &    0.51 $\pm$ 0.30    &    1.4 $\pm$ 0.7    \\
 
Other SM processes  &    0.56 $\pm$ 0.27    &    0.47 $\pm$ 0.23    &    0.63 $\pm$ 0.31    &    0.9 $\pm$ 0.5    &    0.7 $\pm$ 0.4    \\
 
Fake/non-prompt     &    1.1 $\pm$ 0.6    &    1.3 $\pm$ 0.8    &    0.9 $\pm$ 0.6    &    1.3 $\pm$ 1.0    &    1.3 $\pm$ 0.9    \\
 
Charge-flip     &    0.27 $\pm$ 0.14    &    0.11 $\pm$ 0.05    &    0.24 $\pm$ 0.10    &    0.24 $\pm$ 0.11    &    0.13 $\pm$ 0.06    \\
 
\noalign{\smallskip}\hline\hline\noalign{\smallskip}
\end{tabular*} }
\end{center}
\end{table}

In the absence of a significant deviation from the SM prediction,  the results  are interpreted in terms of model-independent upper limits on possible BSM contributions to the SRs, as well as exclusion limits on the masses of the SUSY particles in the benchmark scenarios shown in  Figure~\ref{fig:signals}.
The 95\% CL upper limits on the number of BSM events, $S^{95}$, that may contribute to the SRs are shown in Table~\ref{tab:limits_mi}.
Normalising these by the integrated luminosity $L$ of the data sample, they can be interpreted as upper
limits on the visible BSM cross section ($\sigma_\text{vis}$), defined as $\sigma_\text{vis}= \sigma_\text{prod}\times \mathcal{A}\times \epsilon = S^{95}/L$, where $\sigma_\text{prod}$ is the
production cross section for an arbitrary BSM signal process, and $\mathcal{A}$ and $\epsilon$ are the corresponding acceptance and reconstruction efficiency for the relevant SR. The
probability of the observations being compatible with the SM-only hypothesis is quantified by the $p$-values
displayed in the last column of Table~\ref{tab:limits_mi}.  For SRs where the data yield is smaller than expected, the $p$-value is capped at 0.50.

\begin{table}[h!]
\centering
\caption{
Upper limits at 95\% CL  on the visible cross section
($\sigma_\text{vis}$), on the number of
signal events ($S_{\mathrm{obs}}^{95}$ ), and on the number of signal events given the
expected number of background events and its $\pm 1\sigma$ variations, ($S_{\mathrm{exp}}^{95}$).
The last two columns
indicate the $\text{CL}_\text{b}$  value, i.e.\ the confidence level observed for
the background-only hypothesis, the discovery $p$-value ($p(s = 0)$) and its associated significance $Z$. }
\label{tab:limits_mi}
{\scriptsize
\setlength{\tabcolsep}{0.0pc}
\begin{tabular*}{\textwidth}{@{\extracolsep{\fill}}lccccc}
\noalign{\smallskip}\hline\hline\noalign{\smallskip}{\textbf{SR}}                        & $\sigma_\text{vis}$ [fb]  &  $S_{\mathrm{obs}}^{95}$  & $S_{\mathrm{exp}}^{95}$ & $\text{CL}_\text{b}$ & $p(s=0)$ ($Z$)  \\
\noalign{\smallskip}\hline\hline\noalign{\smallskip}
SRGGWZ-L    & $0.06$ &  $8.3$ & $ { 5.2 }^{ +2.0 }_{ -1.1 }$ & $0.91$ & $ 0.06$~$(1.59)$ \\[1pt]
SRGGWZ-M    & $0.03$ &  $4.8$ & $ { 5.2 }^{ +1.9 }_{ -1.4 }$ & $0.38$ & $ 0.50$~$(0.00)$ \\[1pt]
SRGGWZ-H      & $0.03$ &  $4.2$ & $ { 4.9 }^{ +2.1 }_{ -1.1 }$ & $0.30$ & $ 0.50$~$(0.00)$ \\[1pt]
SRSSWZ-L    & $0.04$ &  $5.9$ & $ { 6.2 }^{ +2.6 }_{ -1.8 }$ & $0.42$ & $ 0.50$~$(0.00)$ \\[1pt]
SRSSWZ-ML    & $0.08$ &  $10.5$ & $ { 6.5 }^{ +2.3 }_{ -1.5 }$ & $0.95$ & $ 0.02$~$(2.07)$ \\[1pt]
SRSSWZ-MH    & $0.06$ &  $8.8$ & $ { 5.5 }^{ +2.1 }_{ -1.4 }$ & $0.93$ & $ 0.04$~$(1.80)$ \\[1pt]
SRSSWZ-H    & $0.06$ &  $8.8$ & $ { 5.5 }^{ +2.3 }_{ -1.4 }$ & $0.91$ & $ 0.10$~$(1.29)$ \\[1pt]
SRGGSlep-L    & $0.03$ &  $4.1$ & $ { 4.8 }^{ +2.0 }_{ -1.1 }$ & $0.34$ & $ 0.50$~$(0.00)$ \\[1pt]
SRGGSlep-M    & $0.04$ &  $6.2$ & $ { 5.7 }^{ +2.4 }_{ -1.4 }$ & $0.60$ & $ 0.42$~$(0.19)$ \\[1pt]
SRGGSlep-H    & $0.02$ &  $2.9$ & $ { 4.5 }^{ +2.0 }_{ -0.9 }$ & $0.01$ & $ 0.40$~$(0.26)$  \\[1pt]
SRSSSlep-L    & $0.08$ &  $11.8$ & $ { 6.0 }^{ +2.3 }_{ -1.8 }$ & $0.98$ & $ 0.01$~$(2.23)$ \\[1pt]
SRSSSlep-ML    & $0.04$ &  $4.9$ & $ { 5.1 }^{ +2.4 }_{ -1.0 }$ & $0.45$ & $ 0.50$~$(0.00)$ \\[1pt]
SRSSSlep-MH    & $0.06$ &  $8.0$ & $ { 5.6 }^{ +2.3 }_{ -1.7 }$ & $0.85$ & $ 0.16$~$(1.01)$ \\[1pt]
SRSSSlep-H    & $0.02$ &  $2.5$ & $ { 3.5 }^{ +1.3 }_{ -0.4 }$ & $0.00$ & $ 0.40$~$(0.25)$ \\[1pt]
SRSSSlep-H (loose)    & $0.07$ &  $10.0$ & $ { 8.2 }^{ +3.5 }_{ -2.2 }$ & $0.70$ & $ 0.31$~$(0.48)$ \\[1pt]
SRLQD    & $0.06$ &  $7.6$ & $ { 5.8 }^{ +2.2 }_{ -1.7 }$ & $0.81$ & $ 0.22$~$(0.76)$ \\[1pt]
SRUDD-1b    & $0.05$ &  $6.9$ & $ { 5.4 }^{ +2.1 }_{ -1.2 }$ & $0.77$ & $ 0.21$~$(0.80)$ \\[1pt]
SRUDD-2b    & $0.05$ &  $6.3$ & $ { 5.2 }^{ +2.5 }_{ -1.2 }$ & $0.70$ & $ 0.26$~$(0.66)$ \\[1pt]
SRUDD-ge2b    & $0.04$ &  $5.9$ & $ { 6.1 }^{ +2.3 }_{ -1.4 }$ & $0.46$ & $ 0.50$~$(0.00)$ \\[1pt]
SRUDD-ge3b    & $0.05$ &  $6.8$ & $ { 6.2 }^{ +2.6 }_{ -1.6 }$ & $0.62$ & $ 0.41$~$(0.22)$ \\%
\noalign{\smallskip}\hline\hline\noalign{\smallskip}
\end{tabular*}}
\end{table}

\begin{figure}[htb!]
 
\begin{center}
 
\begin{subfigure}{0.49\textwidth}
\includegraphics[width=1\textwidth]{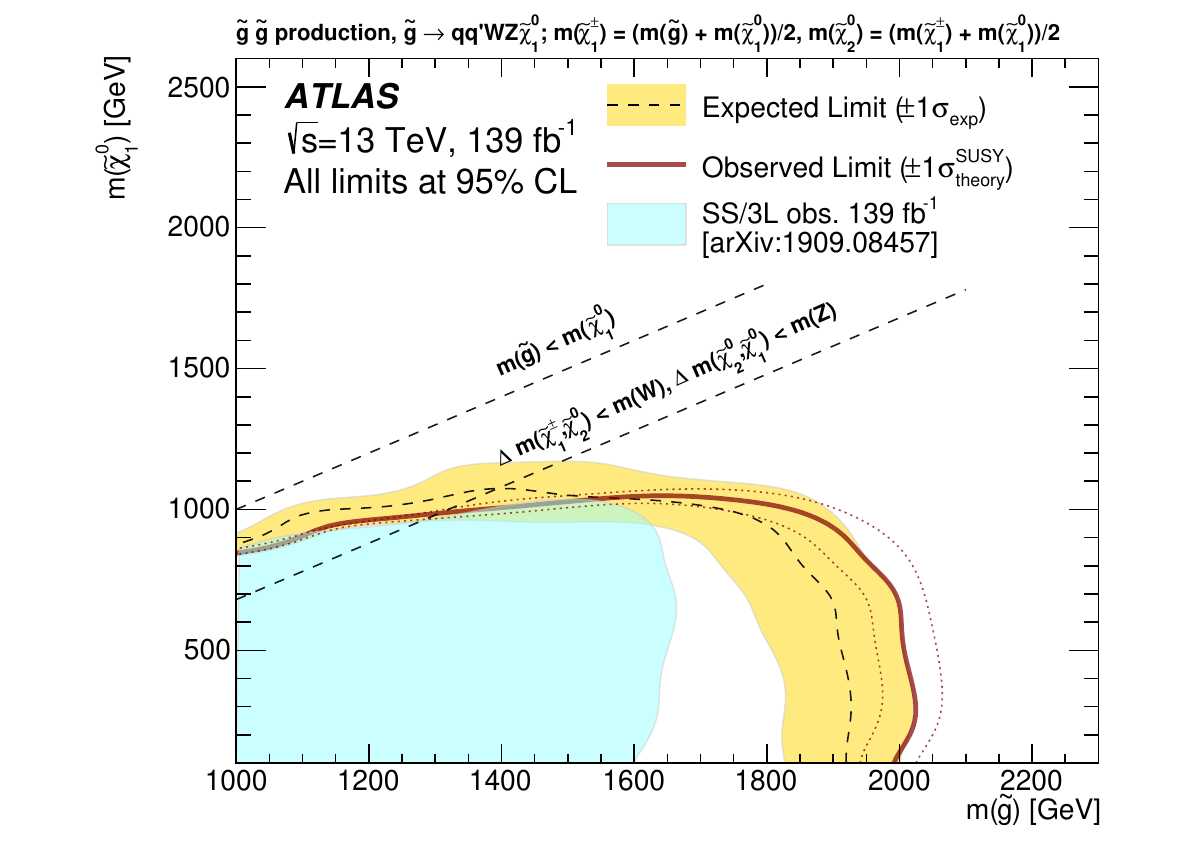}
\caption{$\tilde g \rightarrow qq^\prime WZ \tilde\chi_1^0 $ }
\label{fig:limit_rpcggwz}
\end{subfigure}
\begin{subfigure}{0.49\textwidth}
\includegraphics[width=1\textwidth]{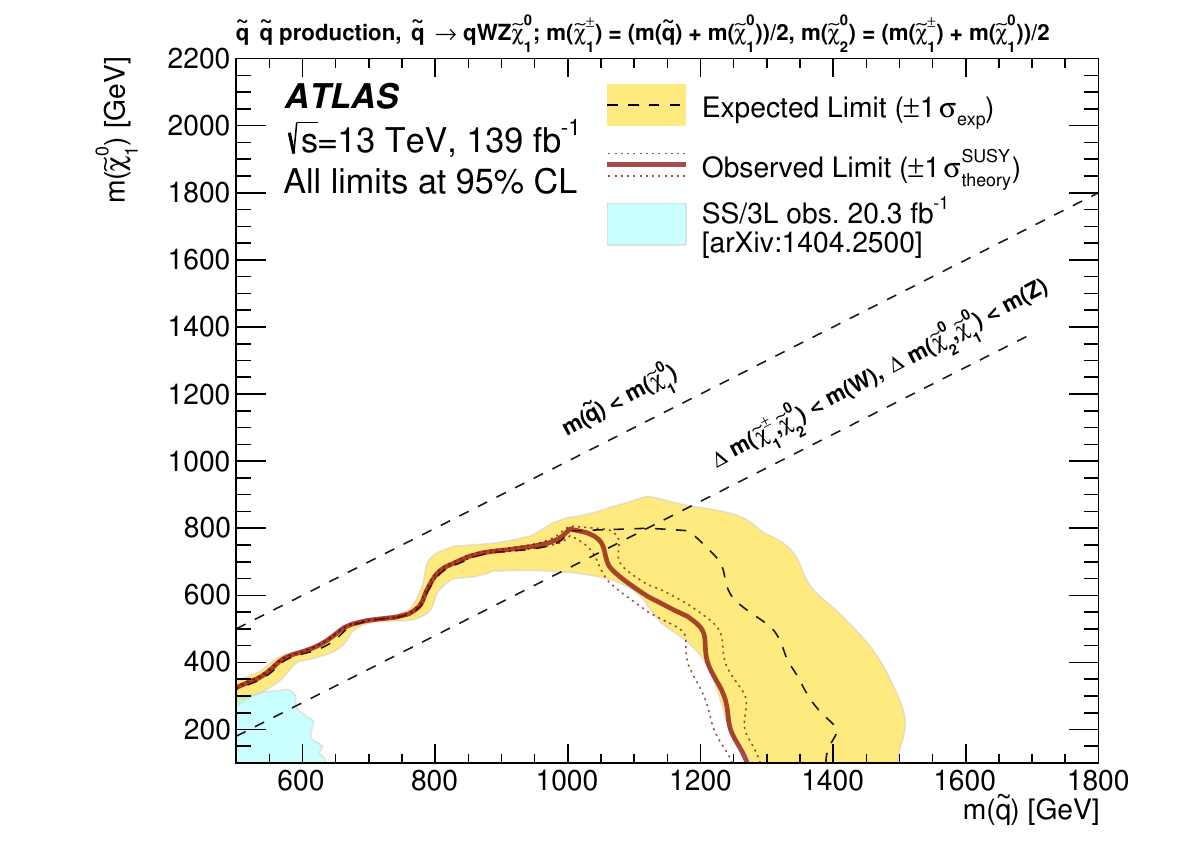}
\caption{$\tilde q \rightarrow q^\prime WZ \tilde\chi_1^0 $  }
\label{fig:limit_rpcsswz}
\end{subfigure}

\begin{subfigure}{0.49\textwidth}
\includegraphics[width=1\textwidth]{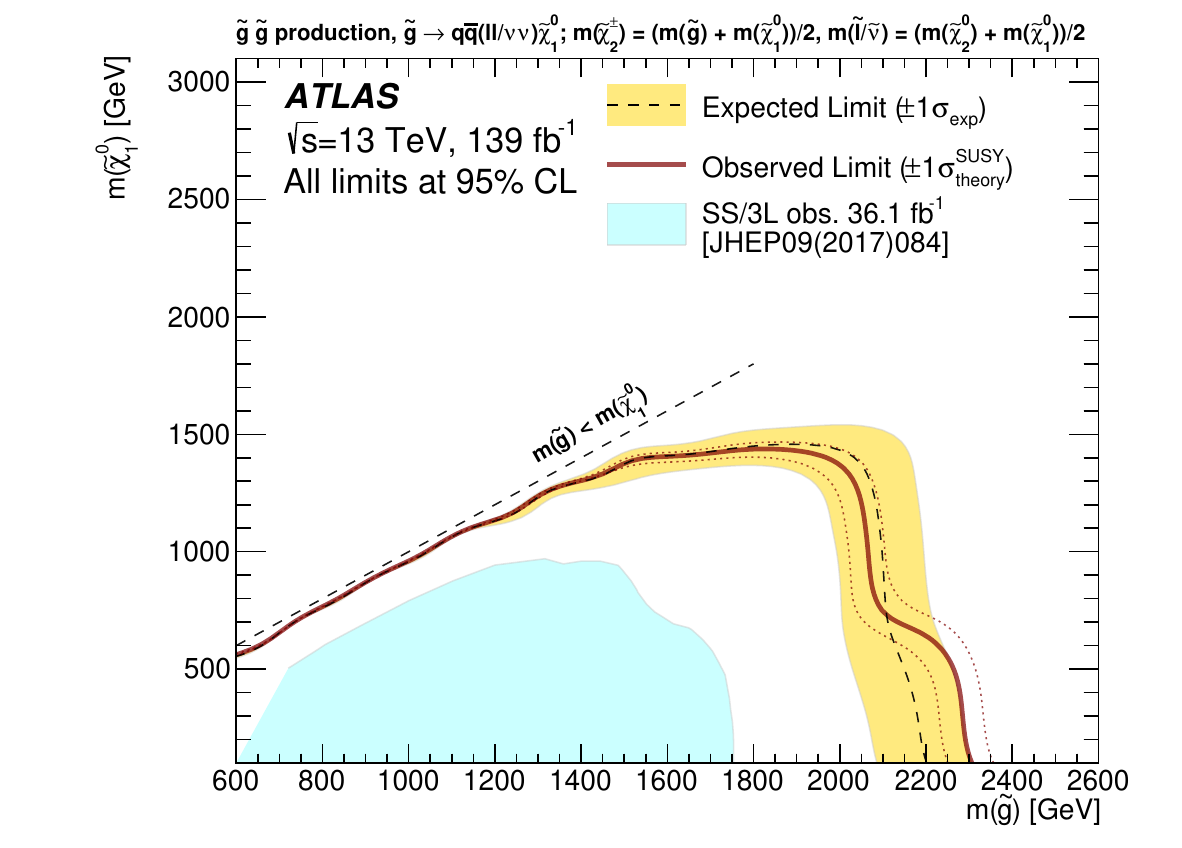}
\caption{ $\tilde g \rightarrow q\bar q(\ell\ell/\nu\nu) \tilde\chi_1^0 $ }
\label{fig:limit_rpcggslep}
\end{subfigure}
\begin{subfigure}{0.49\textwidth}
\includegraphics[width=1\textwidth]{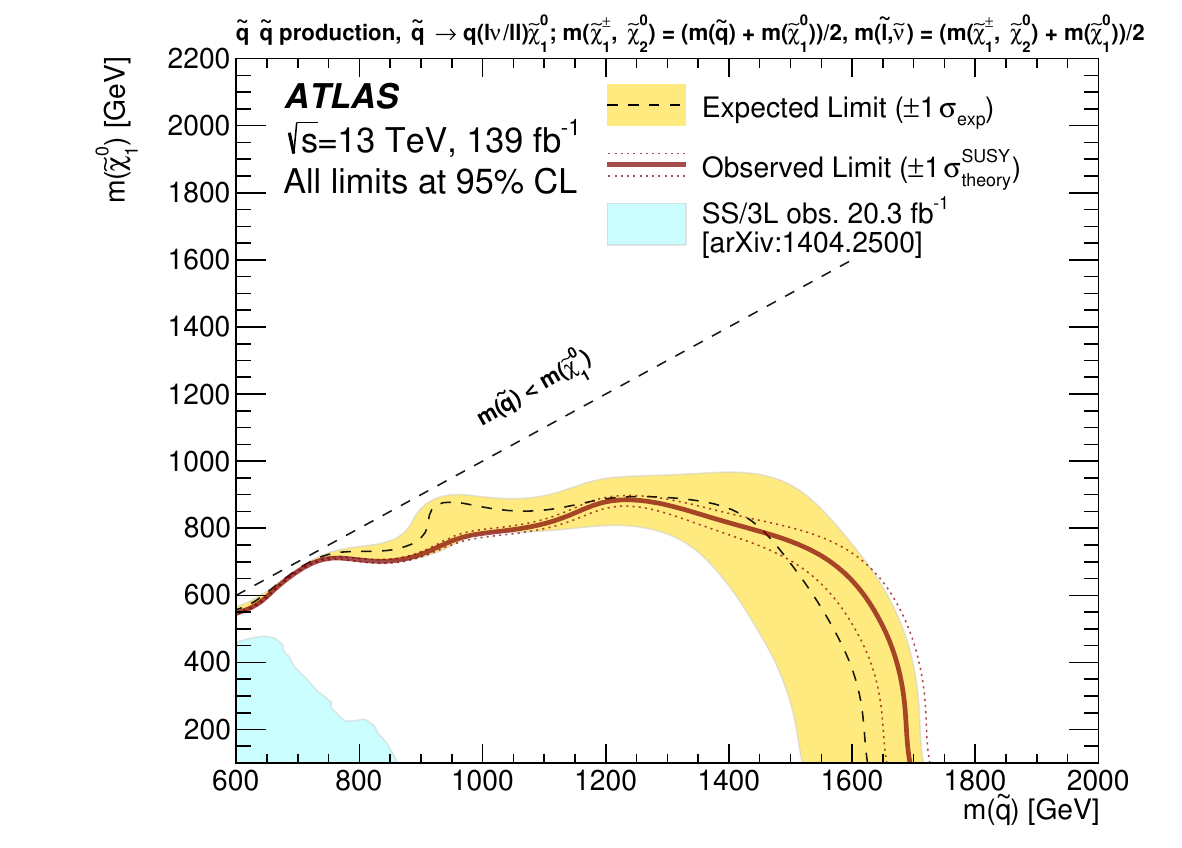}
\caption{$\tilde q \rightarrow   q(\ell\nu/\ell\ell/\nu\nu)\tilde\chi_1^0 $}
\label{fig:limit_rpcssslep}
\end{subfigure}
\end{center}
 
\begin{subfigure}{0.49\textwidth}
\includegraphics[width=1\textwidth]{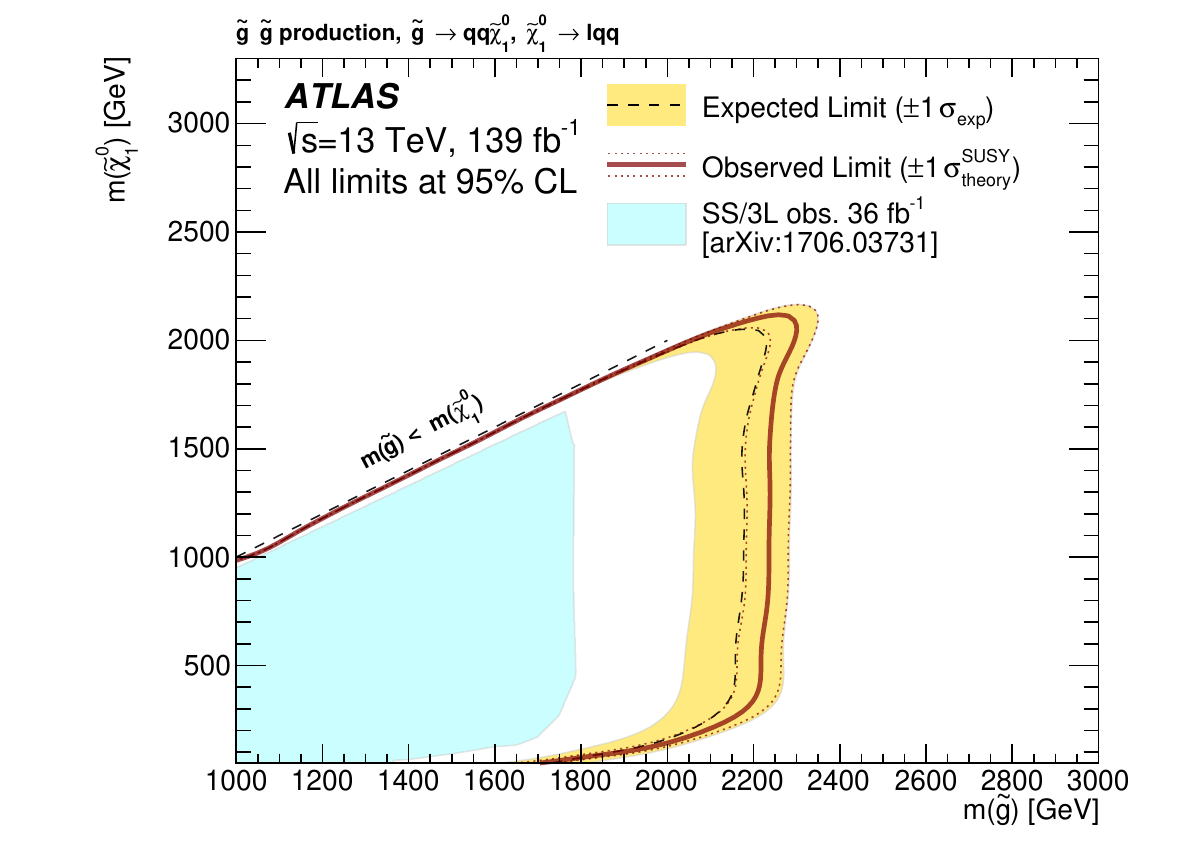}
\caption{$\tilde g \rightarrow q\bar q \tilde\chi_1^0 $, $\tilde\chi_1^0 \rightarrow \ell qq$ }
\label{fig:limit_rpvlqd}
\end{subfigure}
\begin{subfigure}{0.49\textwidth}
\includegraphics[width=1\textwidth]{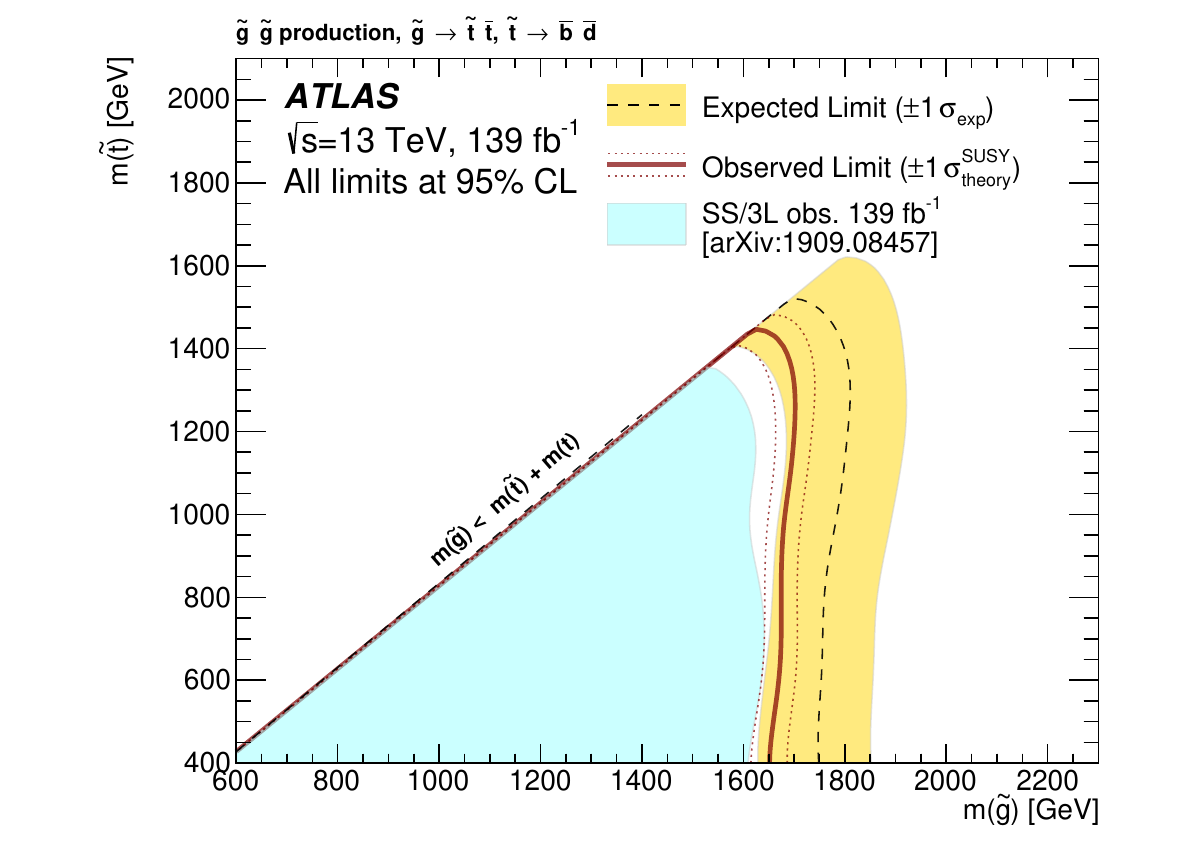}
\caption{ $\tilde g \rightarrow \tilde t \bar{t}$, $\tilde t \rightarrow \bar{b}\bar{d}$}
\label{fig:limit_rpvudd}
\end{subfigure}
 
\caption{ Observed (red line) and expected  (black dashed line) 95\% CL exclusion regions for $\tilde g$, $ \tilde q$, $\tilde\chi_1^0$ and $\tilde t$ masses obtained for the models shown in Figure~\ref{fig:signals}.   The yellow band shows the $ \pm 1 \sigma$ variation of the expected limits.  The red dotted lines around the observed limit illustrate the change in the observed limit as the nominal signal
cross section is scaled up and down by the theoretical uncertainty.  The light blue shaded area  indicates the observed limits obtained by previous ATLAS searches. }
\label{fig:limits_RPC}
\end{figure}

Exclusion limits at 95\% CL are also set on the masses of the superpartners involved in the SUSY benchmark scenarios considered.
Figure~\ref{fig:N1sec10} shows the fitted variable's distribution for each displayed SR.
The bins displayed above the SR requirement correspond to the binning used in the exclusion fit.
For illustration purposes, only one SR per benchmark model is shown.
Figure~\ref{fig:limits_RPC} shows the exclusion limits obtained for the RPC models shown in Figure~\ref{fig:signals}.
For each point of the $\{m_{\tilde g (\tilde q)},m_{\tilde\chi_1^0}\}$ parameter space, the SR with the best expected sensitivity is chosen.
Figures~\ref{fig:limit_rpcggwz} and \ref{fig:limit_rpcsswz} show the mass limits on gluinos and squarks, respectively, for the cascade decays of charginos into pairs of SM bosons.
The mass limits on gluinos  are up to 400~\GeV\ higher than the previously obtained limits and exclude gluinos with masses up to 2~\TeV. 
The improvement over Ref.~\cite{SUSY-2018-09} is achieved mostly by using a looser SR selection
to increase signal acceptance, and by benefiting from the high discriminating power of the \meff shape in SRGGWZ-M.
For squarks,  the mass limits surpass the prior results by around 600~\GeV\ and exclude squarks with masses up to 1.2~\TeV.
 
Figures~\ref{fig:limit_rpcggslep} and \ref{fig:limit_rpcssslep} show the mass limits on gluinos  and squarks, respectively, for the decays of charginos and neutralinos into sleptons.
The mass limits on gluinos extend up to 2.2~\TeV\ for a massless neutralino, while in a very compressed scenario, the limits on the neutralino mass surpass the previous ones by around 200~\GeV.
For the model where squarks are pair produced, the mass limits on squarks are up to 850~\GeV\ higher than the previous limits and exclude squarks with masses up to 1.7~\TeV.
 
Figure~\ref{fig:limit_rpvlqd} shows the mass limits on gluinos for the scenario where the neutralino decays via $\lambda^\prime$ RPV couplings.
Gluinos with masses up to 2.2~\TeV\ are excluded, surpassing the results of the previous search by around 400~\GeV.
The sensitivity decreases rapidly for low-mass neutralinos because their decay products are very collimated and do not pass the isolation requirements, resulting in very low efficiency.
The limits in Figure~\ref{fig:limit_rpvudd} are set for pair production of gluinos in the RPV model where gluinos decay via top squarks into $tbd$ final states if $\lambda''$  couplings are non-zero.
The limits were obtained by using the statistical combination of SRUDD-1b and SRUDD-ge2b, as it was found to provide the best expected sensitivity.
Gluinos with masses up to 1.65~\TeV\ are excluded for a top squark with a mass below 1.45~\TeV.
The improvement over Ref.~\cite{SUSY-2018-09} is in this case driven by the split
into SRs of different $b$-jet multiplicities, together with the binned fits of the \HTJets\ distributions.

\FloatBarrier

\FloatBarrier
 
\section{Conclusion}
\label{sec:conclusion}

A search for pair production of squarks or gluinos decaying via sleptons or weak bosons in final states with either two same-sign leptons or  at least three leptons is presented. The search uses 139~fb$^{-1}$  of proton--proton collision data  collected by the ATLAS detector at the LHC at a centre-of-mass energy of 13~\TeV.

No significant excess of events over the Standard Model expectation is observed.
The results are interpreted in the context of several supersymmetric simplified models featuring gluino or squark pair production in R-parity-conserving and R-parity-violating scenarios.
Lower limits on particle masses are derived at 95\% confidence level for these models, reaching as high as 2.2~\TeV\ for gluinos and 1.7~\TeV\ for squarks, surpassing the exclusion
limits from similar previous searches performed by ATLAS.
Improved analysis techniques, the inclusion of a control region for the $WZ$\,+\,jets background, and a significantly larger data set in some cases, contributed to this improvement.
Model-independent limits on the cross section of a possible signal contribution to the signal regions defined in this search are also set.


\section*{Acknowledgements}


We thank CERN for the very successful operation of the LHC, as well as the
support staff from our institutions without whom ATLAS could not be
operated efficiently.
 
We acknowledge the support of
ANPCyT, Argentina;
YerPhI, Armenia;
ARC, Australia;
BMWFW and FWF, Austria;
ANAS, Azerbaijan;
CNPq and FAPESP, Brazil;
NSERC, NRC and CFI, Canada;
CERN;
ANID, Chile;
CAS, MOST and NSFC, China;
Minciencias, Colombia;
MEYS CR, Czech Republic;
DNRF and DNSRC, Denmark;
IN2P3-CNRS and CEA-DRF/IRFU, France;
SRNSFG, Georgia;
BMBF, HGF and MPG, Germany;
GSRI, Greece;
RGC and Hong Kong SAR, China;
ISF and Benoziyo Center, Israel;
INFN, Italy;
MEXT and JSPS, Japan;
CNRST, Morocco;
NWO, Netherlands;
RCN, Norway;
MEiN, Poland;
FCT, Portugal;
MNE/IFA, Romania;
MESTD, Serbia;
MSSR, Slovakia;
ARRS and MIZ\v{S}, Slovenia;
DSI/NRF, South Africa;
MICINN, Spain;
SRC and Wallenberg Foundation, Sweden;
SERI, SNSF and Cantons of Bern and Geneva, Switzerland;
MOST, Taiwan;
TENMAK, T\"urkiye;
STFC, United Kingdom;
DOE and NSF, United States of America.
In addition, individual groups and members have received support from
BCKDF, CANARIE, Compute Canada and CRC, Canada;
PRIMUS 21/SCI/017 and UNCE SCI/013, Czech Republic;
COST, ERC, ERDF, Horizon 2020 and Marie Sk{\l}odowska-Curie Actions, European Union;
Investissements d'Avenir Labex, Investissements d'Avenir Idex and ANR, France;
DFG and AvH Foundation, Germany;
Herakleitos, Thales and Aristeia programmes co-financed by EU-ESF and the Greek NSRF, Greece;
BSF-NSF and MINERVA, Israel;
Norwegian Financial Mechanism 2014-2021, Norway;
NCN and NAWA, Poland;
La Caixa Banking Foundation, CERCA Programme Generalitat de Catalunya and PROMETEO and GenT Programmes Generalitat Valenciana, Spain;
G\"{o}ran Gustafssons Stiftelse, Sweden;
The Royal Society and Leverhulme Trust, United Kingdom.
 
The crucial computing support from all WLCG partners is acknowledged gratefully, in particular from CERN, the ATLAS Tier-1 facilities at TRIUMF (Canada), NDGF (Denmark, Norway, Sweden), CC-IN2P3 (France), KIT/GridKA (Germany), INFN-CNAF (Italy), NL-T1 (Netherlands), PIC (Spain), ASGC (Taiwan), RAL (UK) and BNL (USA), the Tier-2 facilities worldwide and large non-WLCG resource providers. Major contributors of computing resources are listed in Ref.~\cite{ATL-SOFT-PUB-2021-003}.


\clearpage
\appendix

\printbibliography

\clearpage
 
\begin{flushleft}
\hypersetup{urlcolor=black}
{\Large The ATLAS Collaboration}

\bigskip

\AtlasOrcid[0000-0002-6665-4934]{G.~Aad}$^\textrm{\scriptsize 102}$,
\AtlasOrcid[0000-0002-5888-2734]{B.~Abbott}$^\textrm{\scriptsize 120}$,
\AtlasOrcid[0000-0002-1002-1652]{K.~Abeling}$^\textrm{\scriptsize 55}$,
\AtlasOrcid[0000-0001-5763-2760]{N.J.~Abicht}$^\textrm{\scriptsize 49}$,
\AtlasOrcid[0000-0002-8496-9294]{S.H.~Abidi}$^\textrm{\scriptsize 29}$,
\AtlasOrcid[0000-0002-9987-2292]{A.~Aboulhorma}$^\textrm{\scriptsize 35e}$,
\AtlasOrcid[0000-0001-5329-6640]{H.~Abramowicz}$^\textrm{\scriptsize 151}$,
\AtlasOrcid[0000-0002-1599-2896]{H.~Abreu}$^\textrm{\scriptsize 150}$,
\AtlasOrcid[0000-0003-0403-3697]{Y.~Abulaiti}$^\textrm{\scriptsize 117}$,
\AtlasOrcid[0000-0003-0762-7204]{A.C.~Abusleme~Hoffman}$^\textrm{\scriptsize 137a}$
% ...
\bigskip
\\

% ...
$^{29}$Physics Department, Brookhaven National Laboratory, Upton NY; United States of America.\\
% ...
$^{35}$$^{(a)}$Facult\'e des Sciences Ain Chock, R\'eseau Universitaire de Physique des Hautes Energies - Universit\'e Hassan II, Casablanca;$^{(b)}$Facult\'{e} des Sciences, Universit\'{e} Ibn-Tofail, K\'{e}nitra;$^{(c)}$Facult\'e des Sciences Semlalia, Universit\'e Cadi Ayyad, LPHEA-Marrakech;$^{(d)}$LPMR, Facult\'e des Sciences, Universit\'e Mohamed Premier, Oujda;$^{(e)}$Facult\'e des sciences, Universit\'e Mohammed V, Rabat;$^{(f)}$Institute of Applied Physics, Mohammed VI Polytechnic University, Ben Guerir; Morocco.\\
% ...
$^{49}$Fakult\"{a}t Physik , Technische Universit{\"a}t Dortmund, Dortmund; Germany.\\
% ...
$^{55}$II. Physikalisches Institut, Georg-August-Universit\"{a}t G\"ottingen, G\"ottingen; Germany.\\
% ...
$^{102}$CPPM, Aix-Marseille Universit\'e, CNRS/IN2P3, Marseille; France.\\
% ...
$^{117}$Department of Physics, New York University, New York NY; United States of America.\\
% ...
$^{120}$Homer L. Dodge Department of Physics and Astronomy, University of Oklahoma, Norman OK; United States of America.\\
% ...
$^{137}$$^{(a)}$Departamento de F\'isica, Pontificia Universidad Cat\'olica de Chile, Santiago;$^{(b)}$Millennium Institute for Subatomic physics at high energy frontier (SAPHIR), Santiago;$^{(c)}$Instituto de Investigaci\'on Multidisciplinario en Ciencia y Tecnolog\'ia, y Departamento de F\'isica, Universidad de La Serena;$^{(d)}$Universidad Andres Bello, Department of Physics, Santiago;$^{(e)}$Instituto de Alta Investigaci\'on, Universidad de Tarapac\'a, Arica;$^{(f)}$Departamento de F\'isica, Universidad T\'ecnica Federico Santa Mar\'ia, Valpara\'iso; Chile.\\
% ...
$^{150}$Department of Physics, Technion, Israel Institute of Technology, Haifa; Israel.\\
$^{151}$Raymond and Beverly Sackler School of Physics and Astronomy, Tel Aviv University, Tel Aviv; Israel.\\
% ...
$^{a}$ Also Affiliated with an institute covered by a cooperation agreement with CERN.\\
$^{b}$ Also at An-Najah National University, Nablus; Palestine.\\
$^{c}$ Also at Borough of Manhattan Community College, City University of New York, New York NY; United States of America.\\
$^{d}$ Also at Center for High Energy Physics, Peking University; China.\\
$^{e}$ Also at Center for Interdisciplinary Research and Innovation (CIRI-AUTH), Thessaloniki; Greece.\\
$^{f}$ Also at Centro Studi e Ricerche Enrico Fermi; Italy.\\
% ...

\end{flushleft}


\end{document}